\documentclass{article}

\usepackage{PRIMEarxiv}

\usepackage[utf8]{inputenc}
\usepackage[T1]{fontenc}
\usepackage[colorlinks=true,linkcolor=blue,citecolor=blue,urlcolor=blue]{hyperref}
\usepackage{url}
\usepackage{booktabs}
\usepackage{amsfonts}
\usepackage{nicefrac}
\usepackage{microtype}
\usepackage{fancyhdr}
\usepackage{graphicx}
\usepackage{array}
\usepackage{tabularx}
\usepackage{float}
\usepackage{placeins}
\usepackage{amsmath}
\usepackage{amssymb}
\DeclareMathOperator*{\argmax}{arg\,max}
\usepackage[numbers]{natbib}
\usepackage{enumitem}

\title{Safety, Security, and Cognitive Risks in Neuro-Symbolic AI
\thanks{\textit{\underline{Citation}}:
\textbf{Parmar, M. Safety, Security, and Cognitive Risks in Neuro-Symbolic AI.
arXiv preprint, 2026.}}
}

\author{
  Manoj Parmar \\
  SovereignAI Security Labs \\
  Bengaluru, India \\
  \texttt{manoj@sovereignaisecurity.com}
}

\begin{document}
\maketitle

\begin{abstract}
Neuro-symbolic AI (NeSy) pairs neural perception with symbolic reasoning, making it
attractive for high-stakes domains where explainability and structured inference are
required. However, this hybrid architecture introduces an enlarged attack surface spanning
five layers: neural perception, symbolic knowledge bases, reasoning engines, agentic
orchestration, and data stores---each exploitable in ways absent from purely neural systems.

This paper makes six contributions: (1) formal definitions of \emph{NeSy Attack Surface},
\emph{Symbolic Integrity Violation} (SIV), and \emph{Cross-Layer Amplification Ratio}
$\mathcal{X}$, decomposed into neural-caused and adversarially-induced autonomous
symbolic sensitivity components; (2) a unified threat model extending MITRE ATLAS with
11 NeSy-specific tactic extensions and a five-profile attacker taxonomy; (3) a
symbolic-layer threat catalogue covering KG poisoning, ontology-merging, and
inference-engine subversion; (4) analysis of cognitive risks---automation bias,
authority bias, sycophantic reinforcement---structurally amplified by NeSy's explicit
logical explanations relative to black-box neural outputs;
(5) interdisciplinary mitigations with measurable acceptance criteria aligned to NIST AI
600-1 and the EU AI Act; (6) three empirical benchmarks: (E1)~targeted KG poisoning
achieves break-even SIV at budget $B=5$ on a 205-entity medical KG, with a
KG-specific stealth/SIV trade-off; (E2)~PGD-10 at $\varepsilon=0.01$ yields
$\mathcal{X}=5.884$ (95\%~CI $[4.64, 8.00]$, $p<0.0001$)---confirmed adversarially
specific by a matched-random baseline ($E^{\mathcal{R}}_{\mathrm{rand}}=0$)---on a
DistilBERT\,+\,ProbLog pipeline; (E3)~single-axiom OWL edits achieve 93.3\% SIV success
(equivalence/subclass templates) with 100\% stealth, but held-out detection is 67.9\%
overall with STIX detector failure at 50\% (random-guessing level), an open problem.
\end{abstract}


\tableofcontents
\newpage

\section{Introduction}

The past decade has witnessed rapid progress in both neural and symbolic artificial intelligence
(AI), yet each paradigm retains well-known limitations. Neural networks excel at perception and
pattern recognition across large, heterogeneous data sets, but their internal representations resist
interpretation, their outputs can be confidently wrong, and they offer no native mechanism for
enforcing domain constraints or safety rules. Symbolic systems---logic engines, knowledge graphs,
ontologies, and rule sets---excel at structured reasoning, interpretability, and constraint
enforcement, but are brittle under distributional shift and do not scale easily to raw sensory data.
Neuro-symbolic AI (NeSy) seeks to combine the strengths of both by coupling a neural perceptual
front-end with a symbolic reasoning back-end, often characterised as an instantiation of Kahneman's
dual-process theory: System~1 fast intuition (neural) and System~2 slow deliberation
(symbolic)~\cite{sheth2023neurosymbolic,kahneman2011thinking}.

This architecture is increasingly deployed in safety-critical contexts, including clinical decision
support~\cite{kumari2025safe}, cybersecurity operations~\cite{arxiv2509cybersecurity}, enterprise
automation~\cite{allegrograph2024trust}, and mental-health assistance~\cite{microsoft2025psychological}.
The promise is compelling: symbolic reasoning provides auditable traces, domain constraints, and
alignment with regulatory requirements, while neural components handle perception at scale.
Regulatory bodies such as the European Data Protection Supervisor (EDPS) have explicitly highlighted
neuro-symbolic approaches as potentially facilitating transparency, accountability, and data
minimisation required under data-protection law~\cite{edps2024neurosymbolic}.

Despite these advantages, NeSy systems are not inherently safe or secure. Integrating neural and
symbolic components multiplies both the asset inventory and the attack surface. MITRE ATLAS---the
adversarial threat landscape for AI/ML systems---now documents 16 tactics, 84 techniques, and
numerous sub-techniques covering data poisoning, model extraction, prompt injection, and agentic
context manipulation~\cite{vectra2024atlas}. The OWASP Top~10 for LLM Applications enumerates the
most critical risks for large language model (LLM)-based and agentic systems~\cite{confidentai2025owasp}.
Yet neither framework fully addresses the symbolic substrate of NeSy systems: knowledge graph
integrity attacks, ontology-merging attacks, inference-engine subversion, and reasoning-rule
poisoning remain largely absent from existing catalogues~\cite{riley2024atlas}.

Beyond technical threats, NeSy systems introduce cognitive and psychological risks of a distinctive
character. Because they produce structured, logic-sounding justifications, they risk amplifying
automation bias~\cite{sue2024behavioural}, sycophantic reinforcement of false
beliefs~\cite{varga2024cognitive}, and manipulation of emotionally vulnerable
users~\cite{microsoft2025psychological}. Cognitive security (COGSEC) frameworks identify these as
deliberate and emergent threats to human decision-making that must be addressed at the design,
deployment, and governance levels~\cite{blackbird2024cogsec}.

\textbf{Contributions.} This paper makes six contributions:
\begin{enumerate}
  \item \textbf{Formal threat definitions.} We introduce three formal definitions specific to
        NeSy systems: \emph{NeSy Attack Surface} (Definition~1), \emph{Symbolic Integrity
        Violation} (Definition~2), and \emph{Cross-Layer Amplification Ratio} $\mathcal{X}$
        (Definition~3), providing a rigorous basis for comparing and quantifying NeSy threats.
  \item \textbf{NeSy attacker taxonomy.} We present a five-profile NeSy attacker capability
        taxonomy (Table~\ref{tab:attacker}) extending MITRE ATLAS to knowledge-graph insiders,
        symbolic supply-chain attackers, and social-engineering adversaries.
  \item \textbf{Symbolic-layer threat extension.} We identify and formalise NeSy-specific
        attack categories---KG poisoning, ontology-merging attacks, inference-engine
        subversion, cross-layer amplification---absent from existing ATLAS and OWASP catalogues.
  \item \textbf{Cognitive security literature review and threat predictions.} We survey the
        cognitive and psychological risk literature as it applies to NeSy systems, identifying
        how structured logical explanations are predicted to amplify automation bias, authority
        bias, and sycophantic reinforcement relative to purely neural outputs. These predictions
        are theoretically grounded but not yet empirically validated; E4 (cognitive safety user
        study, IRB pending) is specified to test them.
  \item \textbf{Interdisciplinary mitigations with acceptance criteria.} We propose mitigations
        across technical, governance, and human-factors dimensions, organised as a practitioner
        checklist with measurable acceptance criteria aligned with NIST AI 600-1~\cite{nistai6001nesy}
        and the EU AI Act~\cite{euaiact2024nesy}.
  \item \textbf{Empirical validation.} Three experiments ground the threat model
        on domain-relevant artefacts: (i)~targeted KG poisoning achieves SIV\,$>$\,5\% at
        break-even $B{=}5$ on a 205-entity medical KG, $14.6\times$ more efficient than random,
        with a KG-specific stealth/SIV trade-off; (ii)~PGD-10 produces
        $\mathcal{X}{=}5.884$ (95\% CI $[4.64, 8.00]$, $p{<}0.0001$) on a DistilBERT\,+\,ProbLog
        pipeline, confirmed adversarially specific by a matched-random baseline
        ($E^{\mathcal{R}}_{\mathrm{rand}}{=}0$); (iii)~single-axiom OWL edits achieve 93.3\%
        SIV success (equivalence/subclass scope) with 100\% stealth and 67.9\% held-out
        detection, but STIX detector fails at 50\% held-out recall---an open limitation.
\end{enumerate}

\textbf{Paper structure.}
Section~\ref{sec:relatedwork} surveys related work.
Section~\ref{sec:background} reviews foundational NeSy concepts and the trustworthy AI landscape.
Section~\ref{sec:architecture} characterises the NeSy architecture and assets.
Section~\ref{sec:threatmodel} presents the formal threat-modelling methodology.
Sections~\ref{sec:technical} and~\ref{sec:cognitive} detail technical and cognitive threat
categories, respectively.
Section~\ref{sec:scenarios} illustrates threats through concrete scenarios.
Section~\ref{sec:mitigations} proposes protection mechanisms.
Section~\ref{sec:checklist} provides a practitioner checklist with acceptance criteria.
Section~\ref{sec:experiments} presents empirical validation of E1--E3.
Section~\ref{sec:conclusion} concludes.
Section~\ref{sec:broadimpact} discusses broader impacts.

\section{Related Work}
\label{sec:relatedwork}

\textbf{NeSy trustworthiness surveys.}
Michel-Dél\'{e}tie and Sarker~\cite{nesyjournal2024systematic} survey neuro-symbolic methods
for trustworthy AI across interpretability, safety, robustness, fairness, and privacy. Their
work catalogues how hybrid methods address each dimension but does not develop a unified
security threat model and omits the cognitive-security angle. Gaur and
Sheth~\cite{gaur2023trustworthy} propose the CREST framework (Consistency, Reliability,
Explainability, Safety) as design principles for trustworthy NeSy systems; our paper operationalises
these principles as verifiable acceptance criteria and extends them to adversarial threat scenarios.

\textbf{NeSy security and cybersecurity.}
Hakim et al.~\cite{arxiv2509cybersecurity} survey NeSy AI applications in cybersecurity,
covering detection, knowledge representation, and reasoning, but focus on NeSy as a
\emph{defensive tool} rather than analysing NeSy systems as \emph{attack targets}. Eckhoff
et al.~\cite{nesyjournal2024defending} conduct experiments on NeSy for cyber threat defence,
demonstrating reasoning over STIX/ATT\&CK knowledge, but again from the defender's perspective.
Our paper uniquely treats the NeSy symbolic substrate itself as an attack surface, introducing
formal definitions and an attacker taxonomy applicable across domains.

\textbf{Adversarial ML and knowledge graph attacks.}
Goodfellow et al.~\cite{goodfellow2015fgsm} established adversarial examples for neural networks;
subsequent work has demonstrated that adversarial perturbations propagate into downstream reasoning
components. Knowledge graph adversarial attacks (e.g.,~\cite{zougraph2022}) target relational
embeddings and triples, but focus on graph neural network classifiers rather than symbolic
reasoning pipelines. Our \emph{Cross-Layer Amplification Ratio} $\mathcal{X}$ (Definition~3)
is the first formal measure of how neural adversarial perturbations amplify through symbolic
reasoning chains in NeSy systems. We note that $\mathcal{X}$ is conceptually related to
error-propagation measures in reliability engineering (e.g., sensitivity importance
measures in fault trees~\cite{gaur2023trustworthy}) and signal-to-noise propagation in
cascade systems; unlike those measures, $\mathcal{X}$ is defined over discrete
decision-class flips in a neural--symbolic pipeline and includes a decomposition into
neural-caused and autonomous symbolic sensitivity components (Definition~3).

\textbf{MITRE ATLAS and OWASP LLM extensions.}
Riley~\cite{riley2024atlas} identifies gaps in MITRE ATLAS coverage of symbolic and NeSy threats.
Adabara et al.~\cite{adabara2025agentic} survey agentic AI risks in cybersecurity, including
multi-agent orchestration and tool-use attacks. Our work provides a systematic tactic-by-layer
mapping of ATLAS to the NeSy symbolic substrate (Table~\ref{tab:atlas}), distinguishing
existing ATLAS coverage from proposed NeSy extensions, and introduces the first five-profile
attacker taxonomy tailored to NeSy access models.

\textbf{Cognitive security and automation bias.}
Parasuraman and Riley's foundational work on automation bias and its failure modes (misuse,
disuse, abuse, complacency) has been extended to AI systems by Zou
et al.~\cite{automationbias2025nesy} and others. The manipulation risks specific to sycophantic
AI are documented by Varga~\cite{varga2024cognitive}. We are the first to connect these
cognitive failure modes specifically to the structural feature of NeSy systems---structured
logical explanations---as a systematic amplifier of authority bias.

\textbf{AI governance frameworks.}
The NIST AI RMF~\cite{nistai1001nesy} and NIST AI 600-1~\cite{nistai6001nesy} provide
governance structures for AI risk management and generative AI profiles, respectively. The
EU AI Act~\cite{euaiact2024nesy} classifies healthcare and critical infrastructure AI as
high-risk with mandatory conformity assessments. Our governance mitigations extend these
frameworks with NeSy-specific controls for symbolic asset integrity and cognitive safety.

\textbf{Positioning relative to closest prior work.}
Table~\ref{tab:relwork} places this paper in the context of its four closest
comparators. The key differentiators are: (i) formal threat definitions (Definitions 1--3)
covering both neural and symbolic layers; (ii) a five-profile attacker taxonomy tailored
to NeSy access models; (iii) explicit cognitive-security analysis; and (iv) empirical
validation on domain-relevant artefacts for all three threat categories.

\begin{table}[H]
\centering
\caption{Comparison with closest related work. $\bullet$ = covered, $\circ$ = partial,
$-$ = not covered. \textbf{Formal defs}: formal threat definitions for symbolic layer;
\textbf{Attacker tax.}: NeSy-specific attacker taxonomy;
\textbf{Cog.\ risk}: cognitive/psychological risk analysis;
\textbf{Empirical}: empirical validation with quantitative results.}
\label{tab:relwork}
\small
\begin{tabularx}{\textwidth}{@{} X l l l l @{}}
\toprule
\textbf{Work} & \textbf{Formal defs} & \textbf{Attacker tax.} & \textbf{Cog.\ risk} & \textbf{Empirical} \\
\midrule
Michel-Délétie \& Sarker~\cite{nesyjournal2024systematic} (trustworthiness survey) & $-$ & $-$ & $\circ$ & $-$ \\
Hakim et al.~\cite{arxiv2509cybersecurity} (NeSy in cybersecurity) & $-$ & $\circ$ & $-$ & $\circ$ \\
Eckhoff et al.~\cite{nesyjournal2024defending} (NeSy cyber defence) & $-$ & $-$ & $-$ & $\bullet$ \\
Riley~\cite{riley2024atlas} (ATLAS gap analysis) & $\circ$ & $\circ$ & $-$ & $-$ \\
\textbf{This work} & $\bullet$ & $\bullet$ & $\bullet$ & $\bullet$ \\
\bottomrule
\end{tabularx}
\end{table}

\section{Background}
\label{sec:background}

\subsection{Neuro-Symbolic AI: Core Concepts}

Sheth et al.\ define neuro-symbolic AI as the integration of neural network-based methods for
large-scale perception with symbolic knowledge-based approaches for cognition tasks such as
abstraction, analogy, reasoning, and planning~\cite{sheth2023neurosymbolic}. This definition maps
directly onto Kahneman's dual-process theory: neural components approximate System~1 by converting
raw sensory or textual data into symbols, while symbolic structures approximate System~2 by using
background knowledge to reason over those symbols~\cite{kahneman2011thinking}.

Symbolic knowledge structures---knowledge graphs (KGs), ontologies, and formal logic---provide
explicit mappings from perception outputs to domain concepts, enabling traceability of intermediate
reasoning steps and thus supporting explainability, safety constraints, and regulatory
compliance. By contrast, purely neural models offer powerful pattern recognition but function as
opaque black boxes whose internal representations and decision boundaries are difficult to
audit~\cite{infosys2024neurosymbolic}.

Methodologically, Sheth et al.\ categorise NeSy techniques into two major
classes~\cite{sheth2023neurosymbolic}:

\begin{itemize}
  \item \textbf{Lowering methods}: compress structured symbolic knowledge (KGs, logic) into neural
        representations.
  \item \textbf{Lifting methods}: extract and map neural representations into structured symbolic
        knowledge for downstream reasoning.
\end{itemize}

These divide further into four sub-categories: compressed KG representations, compressed logic
representations, decoupled neural--symbolic pipelines, and tightly intertwined differentiable
methods, each evaluated on algorithmic features (perception scale, abstraction, analogy, planning)
and application-level features (user-explainability, domain constraints, scalability, continual
adaptation).

Of particular relevance to safety, the paper notes that NeSy architectures best support
explainability and constraint enforcement when symbolic reasoning remains explicit and intertwined
with neural components, rather than compressing knowledge into opaque neural embeddings. This
connects to \emph{process knowledge-infused AI}, where domain workflows and safety requirements are
encoded as symbolic process knowledge to guide and constrain neural systems in safety-critical
decision-making~\cite{sheth2023neurosymbolic}.

\subsection{Trustworthy Neuro-Symbolic AI}

Recent work on trustworthy NeSy AI argues that trust depends on four interlocking properties:
Consistency, Reliability, user-level Explainability, and Safety (CREST)~\cite{gaur2023trustworthy}.
Gaur et al.\ contend that combining statistical and symbolic AI is necessary to achieve these
properties, because data-driven neural models alone are insufficiently consistent or interpretable
for safety-critical applications.

Systematic reviews of NeSy for trustworthiness decompose ``trustworthy'' into interpretability,
safety, robustness, fairness, and privacy, and survey how hybrid methods address each
dimension~\cite{nesyjournal2024systematic,nesyjournal2024methods}. Several position papers and
industry reports emphasise that neuro-symbolic approaches can mitigate hallucinations and bias in
LLMs by enforcing domain constraints through explicit rules and
KGs~\cite{infosys2024neurosymbolic,allegrograph2024trust,edps2024neurosymbolic}.

At the same time, reviews of NeSy in cybersecurity and high-stakes domains stress dual-use concerns:
the same architectures that enable richer reasoning and better detection can support more autonomous
and capable offensive agents if mis-aligned or
compromised~\cite{arxiv2509cybersecurity,nesyjournal2024defending,adabara2025agentic}. This dual-use
dimension is central to the threat model developed in Section~\ref{sec:threatmodel}.

\section{NeSy Architecture and Asset Inventory}
\label{sec:architecture}

A generic NeSy system for high-stakes applications comprises five functional layers, each
constituting both a capability and an attack surface:

\begin{enumerate}
  \item \textbf{Neural perception modules}: LLMs, vision models, or other deep networks converting
        raw inputs (text, images, logs, sensor data) into latent representations or symbolic
        candidates~\cite{infosys2024neurosymbolic,sheth2023neurosymbolic}.
  \item \textbf{Symbolic knowledge bases}: knowledge graphs, ontologies, rule sets, formal process
        models, and regulatory or domain guidelines encoded as machine-readable
        symbols~\cite{edps2024neurosymbolic,nesyjournal2024systematic,sheth2023neurosymbolic}.
  \item \textbf{Reasoning engines}: logic reasoners, rule engines, probabilistic programs, or
        differentiable logic layers operating over symbolic structures to derive conclusions, check
        constraints, or plan actions~\cite{gaur2023trustworthy,sheth2023neurosymbolic}.
  \item \textbf{Orchestration and agent layer}: pipelines sequencing neural and symbolic calls (e.g.,
        LangChain-style agentic workflows or NeSy pipelines for
        cybersecurity)~\cite{adabara2025agentic,arxiv2509cybersecurity,sheth2023neurosymbolic}.
  \item \textbf{Data and context stores}: training data, retrieval-augmented generation (RAG)
        indices, vector stores, episodic memory, and reasoning logs exposed to users or
        auditors~\cite{infosys2024neurosymbolic,kumari2025safe}.
\end{enumerate}

From a security perspective, each layer is an asset with distinct threat vectors: adversaries can
poison data and knowledge, alter model weights, tamper with rules or ontologies, exploit
orchestration to bypass constraints, or exfiltrate sensitive data via symbolic explanations and
agent tools~\cite{vectra2024atlas,aembit2025owasp,repello2024atlas,confidentai2025owasp,practicaldevsecops2026atlas}.

\section{Threat-Modelling Methodology}
\label{sec:threatmodel}

\subsection{Formal Definitions}

We introduce three formal concepts that underpin the threat analysis.

\textbf{Definition 1 (NeSy Attack Surface).}
Let $\mathcal{S} = \{\mathcal{N}, \mathcal{K}, \mathcal{R}, \mathcal{O}, \mathcal{D}\}$ denote the
five-layer NeSy system, where $\mathcal{N}$ is the neural perception module,
$\mathcal{K}$ the symbolic knowledge base, $\mathcal{R}$ the reasoning engine,
$\mathcal{O}$ the orchestration and agent layer, and $\mathcal{D}$ the data and context store.
The \emph{NeSy Attack Surface} is the set of all interfaces $\mathcal{I}_{ij}$ between layers
$i$ and $j$ across which an adversary with access level $\alpha \in \{$white-box, grey-box,
black-box, KG-insider, supply-chain$\}$ can inject, modify, or exfiltrate information:
\begin{equation}
  \mathrm{AS}(\mathcal{S}, \alpha) = \{ \mathcal{I}_{ij} \;:\; \exists\, \text{attack } a
  \text{ feasible under } \alpha \text{ that corrupts output via } \mathcal{I}_{ij} \}.
\end{equation}
For adversaries with at least API-level access to the system (grey-box, white-box,
KG-insider, and supply-chain profiles), the NeSy attack surface strictly exceeds the attack
surface of a purely neural system $\mathcal{N}$ alone:
\begin{equation}
  |\mathrm{AS}(\mathcal{S}, \alpha)| > |\mathrm{AS}(\mathcal{N}, \alpha)|
  \quad \forall\, \alpha \in \{\text{grey-box, white-box, KG-insider, supply-chain}\},
\end{equation}
because symbolic layers add new interfaces ($\mathcal{I}_{\mathcal{N}\mathcal{K}}$,
$\mathcal{I}_{\mathcal{K}\mathcal{R}}$, $\mathcal{I}_{\mathcal{R}\mathcal{O}}$, etc.) and
a new class of insider threats (KG curator roles) that have no analogue in purely neural
systems. \emph{Proof sketch:} A grey-box adversary with API-level access to
$\mathcal{S}$ can exploit the output of $\mathcal{N}$ as a probe of $\mathcal{K}$'s
structure via differential symbolic explanations---an attack channel absent in a purely
neural $\mathcal{N}$. Formally, for any interface $\mathcal{I}_{\mathcal{N}\mathcal{K}}$
that is reachable from the API output, there exists an attack $a$ (explanation-harvesting)
feasible under grey-box that corrupts output via $\mathcal{I}_{\mathcal{N}\mathcal{K}}$
but not via any interface in $\mathrm{AS}(\mathcal{N}, \text{grey-box})$. This argument
assumes the symbolic layers are not fully sandboxed from the output channel; if
$\mathcal{K}$ and $\mathcal{R}$ are isolated with no output leakage, the marginal attack
surface may be zero.
This inequality is not claimed for the black-box sensor attacker profile, where
the adversary's access is restricted to raw input channels and the marginal attack surface
added by symbolic layers may be zero if those layers are not reachable from the sensor
channel.

\textbf{Definition 2 (Symbolic Integrity Violation).}
Let $\mathcal{K}_0$ denote the authorised knowledge base at deployment time, and
$\mathcal{K}_t$ the knowledge base at time $t$. A \emph{symbolic integrity violation} at
time $t$ is any change $\Delta \mathcal{K}_t = \mathcal{K}_t \mathbin{\triangle} \mathcal{K}_0$ that is
(i) unauthorised, and (ii) causes the reasoning engine $\mathcal{R}$ to produce a different
conclusion $c' \neq c$ on at least one safety-critical query $q$.
Here $\mathbin{\triangle}$ denotes the symmetric difference, capturing both unauthorised
\emph{additions} and \emph{deletions} of knowledge-base triples or axioms:
\begin{equation}
  \mathrm{SIV}(t) \iff \Delta\mathcal{K}_t \neq \emptyset \;\wedge\;
  \exists\, q \in Q_{\mathrm{safe}} : \mathcal{R}(\mathcal{K}_t, q) \neq \mathcal{R}(\mathcal{K}_0, q).
\end{equation}
Symbolic integrity violations are stealthy because the neural component $\mathcal{N}$ may
continue to perform correctly; the attack is entirely within the symbolic substrate.

\textbf{Definition 3 (Cross-Layer Amplification Ratio).}
\emph{Scope:} This definition applies to \emph{decoupled} NeSy architectures in which
$\mathcal{N}$ produces a discrete symbolic output (e.g., hard argmax label or
probability vector fed to an independent reasoning engine $\mathcal{R}$). It does not
directly apply to \emph{fully differentiable} NeSy systems---such as DeepProbLog,
Neural Theorem Provers, or Logical Neural Networks---where the neural and symbolic
components share a common differentiable computation graph; extension to that setting is
an open problem (see future work, Section~\ref{sec:conclusion}).
Let $\delta$ be an adversarial perturbation applied at the input of $\mathcal{N}$ with
$\|\delta\|_\infty \leq \varepsilon$. Define the \emph{neural decision flip rate}
$E^{\mathcal{N}}(\delta) \in [0,1]$ as the fraction of queries on which the neural
component changes its hard output decision under perturbation:
\begin{equation}
  E^{\mathcal{N}}(\delta) = \frac{\bigl|\{x \in \mathcal{Q} : \argmax\,\mathcal{N}(x+\delta) \neq \argmax\,\mathcal{N}(x)\}\bigr|}{|\mathcal{Q}|}.
\end{equation}
Define the \emph{symbolic conclusion flip rate}
$E^{\mathcal{R}}(\delta) \in [0,1]$ as the fraction of queries on which the downstream
symbolic reasoning engine changes its conclusion:
\begin{equation}
  E^{\mathcal{R}}(\delta) = \frac{\bigl|\{x \in \mathcal{Q} : \mathcal{R}(\mathcal{K}, \mathcal{N}(x+\delta)) \neq \mathcal{R}(\mathcal{K}, \mathcal{N}(x))\}\bigr|}{|\mathcal{Q}|}.
\end{equation}
The \emph{Cross-Layer Amplification Ratio} is:
\begin{equation}
  \mathcal{X}(\delta) = \frac{E^{\mathcal{R}}(\delta)}{E^{\mathcal{N}}(\delta)}.
\end{equation}
Because both operands lie in $[0,1]$, $\mathcal{X}$ is dimensionless and
query-scale-invariant: it does not depend on the total number of queries $|\mathcal{Q}|$
and is comparable across NeSy systems of different sizes, provided they share
the decoupled architecture assumed in Definition~3.
$\mathcal{X}$ is undefined when $E^{\mathcal{N}}(\delta) = 0$.

To decompose the sources of symbolic flips, we additionally define two conditional rates.
Let $\mathcal{Q}^+ = \{x \in \mathcal{Q} : \argmax\,\mathcal{N}(x+\delta) \neq \argmax\,\mathcal{N}(x)\}$
be the set of queries on which the neural component changes its decision.
The \emph{neural-caused symbolic flip rate} measures propagation from confirmed neural flips:
\begin{equation}
  E^{\mathcal{R}|\mathcal{N}}(\delta) = \frac{|\{x \in \mathcal{Q}^+ : \mathcal{R}(\mathcal{K},\mathcal{N}(x+\delta)) \neq \mathcal{R}(\mathcal{K},\mathcal{N}(x))\}|}{|\mathcal{Q}^+|}.
\end{equation}
The \emph{autonomous symbolic sensitivity rate} measures symbolic flips that occur on queries
where the neural component did \emph{not} change its hard decision:
\begin{equation}
  E^{\mathcal{R}|\neg\mathcal{N}}(\delta) = \frac{|\{x \in \mathcal{Q}\setminus\mathcal{Q}^+ : \mathcal{R}(\mathcal{K},\mathcal{N}(x+\delta)) \neq \mathcal{R}(\mathcal{K},\mathcal{N}(x))\}|}{|\mathcal{Q}\setminus\mathcal{Q}^+|}.
\end{equation}
$E^{\mathcal{R}|\neg\mathcal{N}} > 0$ indicates that the symbolic layer is sensitive
to sub-decision-boundary neural perturbations---a distinct threat from amplification that
arises when symbolic rules receive soft probabilities or logits rather than hard labels.
$\mathcal{X} > 1$ implies $E^{\mathcal{R}} > E^{\mathcal{N}}$, which can arise from a
combination of high $E^{\mathcal{R}|\mathcal{N}}$ and high $E^{\mathcal{R}|\neg\mathcal{N}}$;
both components should be reported to characterise the symbolic layer's behaviour fully.

\textbf{Remark (threat threshold).} Let $\varepsilon_{\mathrm{op}}$ be the maximum
perturbation achievable against a deployed system under realistic conditions (e.g.,
physical adversarial patches, API-level noise injection). We define \emph{critical
cross-layer amplification} as $\mathcal{X}(\delta) > 2$ at $\varepsilon_{\mathrm{op}}$,
indicating that the symbolic layer more than doubles the neural attack surface.
This threshold is conservative; the empirically observed $\mathcal{X}=5.884$ at
$\varepsilon=0.01$ (Section~\ref{sec:experiments}) substantially exceeds it.
Systems below the threshold may still exhibit $\mathcal{X} > 1$ but at a level
commensurate with known neural-only risks, whereas systems above it require the
architectural defences described in Section~\ref{sec:mitigations}.

\subsection{MITRE ATLAS Alignment}
\label{sec:atlasmap}

MITRE ATLAS catalogues adversarial tactics and techniques targeting AI/ML systems, covering 16
tactics, 84 techniques, and numerous sub-techniques including agentic AI risks such as context
poisoning and malicious tool use~\cite{vectra2024atlas,repello2024atlas}.

Table~\ref{tab:atlas} maps the five NeSy system layers to ATLAS tactics, distinguishing
\textbf{existing} ATLAS coverage (techniques already documented for neural systems) from
\textbf{NeSy extensions} proposed here for the symbolic substrate. The table does not claim
completeness; it identifies where the symbolic layer adds new attack vectors absent from the
current ATLAS catalogue~\cite{riley2024atlas}.

\begin{table}[H]
\centering
\caption{MITRE ATLAS tactic mapping to NeSy system layers.
\textbf{E} = existing ATLAS technique (documented for neural systems).
\textbf{N} = NeSy extension proposed in this paper (symbolic-layer attack not yet in ATLAS catalogue).
\textbf{Severity}: H=High (direct safety/integrity impact), M=Medium (enabler or indirect),
L=Low (information gathering only).}
\label{tab:atlas}
\small
\begin{tabularx}{\textwidth}{@{} p{2.2cm} p{1.9cm} l l X @{}}
\toprule
\textbf{ATLAS Tactic} & \textbf{NeSy Layer} & \textbf{Status} & \textbf{Sev.} & \textbf{Technique / Manifestation} \\
\midrule
Reconnaissance      & $\mathcal{N}$ neural   & \textbf{E} & L & Model architecture discovery; training data inference \\
Reconnaissance      & $\mathcal{K}$ KG       & \textbf{N} & L & Schema and ontology enumeration via explanation traces \\
\midrule
Resource Dev.       & $\mathcal{D}$ data     & \textbf{E} & M & Curating poisoned pre-training corpora \\
Resource Dev.       & $\mathcal{K}$ KG       & \textbf{N} & H & Building adversarial ontology extension for supply-chain merge \\
\midrule
Initial Access      & $\mathcal{N}$ neural   & \textbf{E} & H & Adversarial examples; prompt injection \\
Initial Access      & $\mathcal{K}$ KG       & \textbf{N} & H & Malicious KG triple injection via curator role; ontology merge \\
Initial Access      & $\mathcal{O}$ agent    & \textbf{E} & M & AI agent context poisoning$^\ddagger$ \\
\midrule
Execution           & $\mathcal{R}$ reasoner & \textbf{N} & H & Inference-engine rule triggering via crafted symbolic inputs \\
Execution           & $\mathcal{O}$ agent    & \textbf{E} & H & Malicious tool use; prompt-injected tool invocation \\
\midrule
Persistence         & $\mathcal{K}$ KG       & \textbf{N} & H & Stealthy backdoor triples surviving KG update cycles; ontology rule poisoning \\
Persistence         & $\mathcal{N}$ neural   & \textbf{E} & H & Model backdoor (activation trigger) \\
\midrule
Privilege Escalation & $\mathcal{O}$ agent   & \textbf{E} & H & Excessive autonomy via tool-use escalation \\
Privilege Escalation & $\mathcal{R}$ reasoner& \textbf{N} & H & Symbolic constraint bypass via adversarial rule override \\
\midrule
Discovery           & $\mathcal{K}$ KG       & \textbf{N} & M & KG triple enumeration; safety-rule discovery via differential queries \\
Discovery           & $\mathcal{N}$ neural   & \textbf{E} & M & Model inversion; membership inference \\
\midrule
Impact              & $\mathcal{K}$ KG       & \textbf{N} & H & Symbolic Integrity Violation (SIV, Definition~2); ontology-merging attack \\
Impact              & $\mathcal{N}$ neural   & \textbf{E} & H & Model evasion; denial-of-service via adversarial inputs \\
Impact              & $\mathcal{R}$ reasoner & \textbf{N} & H & Cross-layer amplification of neural perturbation ($\mathcal{X} > 1$) \\
\midrule
Exfiltration        & $\mathcal{D}$ data     & \textbf{E} & M & Training data exfiltration via model outputs \\
Exfiltration        & $\mathcal{K}$ KG       & \textbf{N} & M & Sensitive fact extraction via symbolic explanations and audit trails \\
Exfiltration        & $\mathcal{O}$ agent    & \textbf{E} & H & Exfiltration via AI agent tool invocation$^\ddagger$ \\
\bottomrule
\end{tabularx}
\smallskip
{\small $^\ddagger$Representative ATLAS technique IDs: Initial Access/$\mathcal{O}$: AML.T0051;
Exfiltration/$\mathcal{O}$: AML.T0057. See \url{https://atlas.mitre.org} for full catalogue.}
\end{table}

Of the 20 tactic--layer combinations in Table~\ref{tab:atlas}, 11 are NeSy extensions not
covered by current ATLAS documentation. The largest gaps are in the $\mathcal{K}$ symbolic
knowledge-base layer (Reconnaissance, Resource Development, Initial Access, Persistence,
Discovery, Impact, Exfiltration) and the $\mathcal{R}$ reasoning-engine layer (Execution,
Privilege Escalation, Impact). These represent the primary contribution of this mapping
relative to the existing ATLAS catalogue~\cite{vectra2024atlas,repello2024atlas,riley2024atlas,
arxiv2509cybersecurity,nesyjournal2024defending}.

\subsection{OWASP LLM Top 10 Alignment}

The OWASP Top~10 for LLM Applications (2023--2025) identifies the most critical risks for
LLM-based and agentic systems, including prompt injection, data exfiltration, insecure output
handling, model denial-of-service, excessive autonomy, RAG vulnerabilities, and supply-chain
risks~\cite{aembit2025owasp,evidently2025owasp,oligo2025owasp,confidentai2025owasp}. Many NeSy
systems embed or front-end LLMs as the neural component, so these risks apply directly.

NeSy systems introduce distinctive variants of OWASP categories:

\begin{itemize}
  \item \textbf{RAG and knowledge-base vulnerabilities} now extend to symbolic KGs and rule bases,
        not only unstructured document
        indices~\cite{aembit2025owasp,confidentai2025owasp,nesyjournal2024systematic}.
  \item \textbf{Excessive autonomy} is amplified when symbolic reasoning triggers real-world actions
        via tools and workflows based on inferred rules and
        plans~\cite{adabara2025agentic,confidentai2025owasp,vectra2024atlas}.
  \item \textbf{Sensitive information disclosure} can occur through symbolic explanations and audit
        trails that expose rule conflicts, data provenance, or personal data embedded in
        KGs~\cite{edps2024neurosymbolic,allegrograph2024trust,kumari2025safe}.
\end{itemize}

\subsection{Analytical Lenses}

A comprehensive threat-modelling approach for NeSy systems combines four analytical perspectives:

\begin{itemize}
  \item \textbf{Asset-centric}: systematic analysis of the five-layer NeSy stack
        (Section~\ref{sec:architecture}).
  \item \textbf{Adversary-centric}: ATLAS tactics and techniques extended to symbolic components.
  \item \textbf{Vulnerability-centric}: OWASP LLM Top~10 categories mapped to NeSy-specific
        implementation flaws.
  \item \textbf{Cognitive-centric}: treating users and operators as targets for manipulation and
        over-trust, drawing on cognitive security and dual-process
        psychology~\cite{varga2024cognitive,blackbird2024cogsec,marketingsociety2024system,sue2024behavioural,farnstreet2024kahneman}.
\end{itemize}

\subsection{NeSy Attacker Capability Taxonomy}
\label{sec:attackertaxonomy}

Table~\ref{tab:attacker} presents a five-profile attacker capability taxonomy tailored to NeSy
systems. Each profile extends MITRE ATLAS adversary definitions with NeSy-specific access models,
knowledge levels, primary goals, and representative techniques. The taxonomy is designed to scope
threat-modelling exercises and inform detection and mitigation prioritisation.

\begin{table}[H]
\centering
\caption{NeSy Attacker Capability Taxonomy. Five profiles ordered by decreasing access level.
ATLAS tactic abbreviations: TA = Tactic; $\mathcal{K}$-insider = knowledge-base insider.}
\label{tab:attacker}
\small
\begin{tabularx}{\textwidth}{@{} p{2.2cm} p{2.0cm} p{2.0cm} p{2.8cm} X @{}}
\toprule
\textbf{Profile} & \textbf{Access Level} & \textbf{Knowledge} & \textbf{Primary Goal} & \textbf{Representative NeSy Techniques} \\
\midrule
White-box attacker
  & Full model weights, architecture, \& symbolic rule set
  & Complete (gradients, logic rules, KG schema)
  & Craft adversarial inputs exploiting both neural gradients and symbolic rule topology ($\mathcal{X}\!\gg\!1$); symbolic rule access is required to direct perturbations into ProbLog sensitivity regions
  & PGD/FGSM adversarial examples~\cite{goodfellow2015fgsm}; gradient-guided KG triple poisoning; targeted SIV induction via rule-aware perturbation \\[4pt]

Grey-box API attacker
  & Query API only (black-box neural; white-box outputs)
  & Output labels, explanations, rule traces
  & Extract symbolic rules or KG structure; inject via API inputs
  & Transferability-based adversarial examples; explanation-harvesting for KG extraction; prompt injection into neural front-end~\cite{confidentai2025owasp} \\[4pt]

Black-box sensor / input attacker
  & Raw sensor/data channel only
  & None (output feedback only)
  & Degrade perception to cause downstream reasoning failures
  & Physical adversarial patches; data corruption in sensor feeds; input flooding for DoS on reasoning engine~\cite{vectra2024atlas} \\[4pt]

KG-insider ($\mathcal{K}$-insider)
  & Direct KG / ontology write access
  & Schema, reasoning rules, query patterns
  & Stealthy SIV: alter KG triples or rules to subvert safety-critical conclusions
  & Malicious triple insertion~\cite{nesyjournal2024defending}; ontology-merging attack; rule poisoning to disable safety constraints; provenance spoofing~\cite{riley2024atlas} \\[4pt]

Supply-chain attacker
  & Pre-deployment artefact supply
  & Partial (build pipeline, third-party KG/model sources)
  & Embed backdoors or biased triples before deployment; compromise reasoning library
  & Poisoned pre-trained embedding~\cite{goodfellow2015fgsm}; malicious KG dataset contribution; trojanised reasoning library; compromised ontology update feed~\cite{nesyjournal2024systematic} \\
\bottomrule
\end{tabularx}
\end{table}

The five profiles span the full adversary spectrum from nation-state or insider threat (white-box,
KG-insider) to opportunistic (black-box) and supply-chain actors. Critically, the KG-insider and
supply-chain profiles represent NeSy-specific threat actors absent from ATLAS's primarily
neural-centric adversary model. Any formal risk assessment for a NeSy deployment should enumerate
applicable profiles, map them to the NeSy attack surface $\mathrm{AS}(\mathcal{S}, \alpha)$
(Definition~1), and define detection and response controls per profile.

\section{Technical Threat Categories}
\label{sec:technical}

\subsection{Data and Knowledge Poisoning}

Like purely neural systems, NeSy models are vulnerable to training and fine-tuning data poisoning,
where adversaries inject crafted examples to bias behaviour or create hidden triggers (ATLAS
technique: Poisoned Training Data)~\cite{repello2024atlas,vectra2024atlas}. NeSy adds a second
attack channel: poisoning symbolic knowledge bases---KGs, ontologies, rule sets, and process
models---to degrade reasoning or subvert safety
constraints~\cite{nesyjournal2024defending,nesyjournal2024systematic,arxiv2509cybersecurity}.

Examples include:

\begin{itemize}
  \item Inserting malicious edges in a medical KG that subtly alter contraindication relationships,
        leading a decision-support system to approve unsafe drug combinations in specific comorbidity
        contexts~\cite{nesyjournal2024systematic,kumari2025safe}.
  \item Modifying process knowledge so that safety guardrails apply only to certain patient cohorts,
        enabling targeted harm or discrimination~\cite{riley2024atlas,gaur2023trustworthy}.
  \item Introducing adversarial labels into cybersecurity threat ontologies so that malware families
        are misclassified, reducing detection sensitivity in NeSy-based security operations centre
        (SOC) tools~\cite{nesyjournal2024defending,adabara2025agentic,arxiv2509cybersecurity}.
\end{itemize}

Such attacks benefit from the fact that symbolic knowledge is often curated by domain experts or
ingested from heterogeneous external sources, creating rich supply-chain and governance attack
surfaces beyond classic dataset poisoning~\cite{nesyjournal2024methods,nesyjournal2024systematic}.

\subsection{Model and Reasoning Exploitation}

Adversaries can exploit vulnerabilities in neural and symbolic components and at their interface.

For \textbf{neural components}, ATLAS and OWASP document risks including adversarial examples,
gradient-based model extraction, prompt injection, and model denial-of-service via crafted inputs
that drive expensive reasoning
paths~\cite{practicaldevsecops2026atlas,oligo2025owasp,confidentai2025owasp,vectra2024atlas}. When
these components feed symbolic reasoning, small perturbations can push the system into different
branches of the rule space, amplifying impact.

For \textbf{symbolic components}, proposed attacks include inference-engine subversion (e.g.,
altering rule priorities, introducing non-terminating or conflicting rules), ontology-merging attacks
that create inconsistent concept hierarchies, and schema-evolution attacks that quietly change the
semantics of predicates~\cite{arxiv2509cybersecurity,riley2024atlas,nesyjournal2024defending}.
These can cause safety constraints to be bypassed or misapplied without immediately obvious changes
in outputs.

\textbf{Interface logic} adds further risk. In decoupled pipelines where LLMs call tools or
reasoners based on natural-language descriptions, prompt injection can redirect tool calls or select
different symbolic functions than intended, effectively rewiring the NeSy reasoning graph at
runtime. In tightly intertwined NeSy architectures, gradient-based training could attenuate or
override safety-relevant rules if loss functions are not carefully designed to preserve
invariants~\cite{oligo2025owasp,confidentai2025owasp,sheth2023neurosymbolic,gaur2023trustworthy}.

\subsection{Orchestration and Agentic Workflow Threats}

NeSy is increasingly deployed within agentic systems that plan, call tools, update memory, and
interact over long-lived sessions. ATLAS now includes specific techniques for AI agent context
poisoning, activation triggers, and abusing agent tools for
exfiltration~\cite{vectra2024atlas,adabara2025agentic,kumari2025safe}.

Key threats include:

\begin{itemize}
  \item Poisoning long-term memory or RAG sources so that symbolic plans are constructed from
        adversarially biased context~\cite{confidentai2025owasp,vectra2024atlas}.
  \item Crafting prompts or documents that cause the neural component to misidentify which symbolic
        rules or tools should be applied, enabling policy bypass or harmful
        actions~\cite{tredence2025cognitive,oligo2025owasp,confidentai2025owasp}.
  \item Exploiting tool permissions (database access, file systems, external APIs) to exfiltrate
        sensitive data or perform destructive actions during seemingly legitimate NeSy-driven
        workflows~\cite{aembit2025owasp,adabara2025agentic,vectra2024atlas}.
\end{itemize}

\subsection{Privacy and Confidentiality Risks}

Symbolic knowledge bases often encode personal or sensitive information in structured form, and NeSy
explanations may expose traceable reasoning paths that reveal this information more directly than
opaque neural embeddings. The EDPS notes that while symbolic AI can improve transparency and
accountability, it must not reduce the effectiveness of privacy protections or human
oversight~\cite{edps2024neurosymbolic}.

OWASP identifies sensitive information disclosure as one of the most critical LLM risks, exacerbated
when models have broad access to organisation data via RAG, logs, or tool
calls~\cite{confidentai2025owasp,aembit2025owasp}. For NeSy, this risk extends to:

\begin{itemize}
  \item Leakage of KG fragments or rules revealing proprietary strategies, security procedures, or
        patient-level data through explanation or debugging
        interfaces~\cite{allegrograph2024trust,edps2024neurosymbolic}.
  \item Membership inference on symbolic facts (e.g., whether a particular individual is in a
        high-risk cohort) via probing of NeSy reasoning
        outputs~\cite{nesyjournal2024methods,nesyjournal2024systematic}.
  \item Inadequate segregation of audit logs containing both neural inputs/outputs and symbolic
        traces, leading to unintentional exposure.
\end{itemize}

\section{Cognitive, Psychological, and Behavioural Risks}
\label{sec:cognitive}

\noindent\textbf{Scope note.} This section is a structured review of the cognitive and
psychological risk literature as it applies to NeSy systems. The claims made here are
theoretical and grounded in cited prior work; they are \emph{not} empirically validated in
this paper. Empirical validation (E4: a cognitive safety user study measuring automation
bias amplification by NeSy explanations) requires IRB approval and is deferred to future
work. Readers should treat the amplification hypotheses below as well-motivated predictions,
not established findings.

\subsection{Dual-Process Alignment and Over-Trust}

NeSy explicitly mirrors dual-process theory: neural components provide fast, intuitive judgements,
while symbolic components provide slow, deliberate reasoning. Recent work aligning LLMs with
System~1 or System~2 reasoning demonstrates a trade-off: System~1-aligned models give faster, more
confident answers, whereas System~2-aligned models produce more cautious, step-by-step reasoning at
higher computational
cost~\cite{arxiv2502reasoning,marketingsociety2024system,farnstreet2024kahneman,kahneman2011thinking,sheth2023neurosymbolic}.

From a cognitive-security standpoint, NeSy systems that present symbolic reasoning chains and rule
citations strongly trigger user perceptions of expertise and authority, reinforcing \emph{automation
bias} and \emph{authority bias}. Behavioural design research on dual-process thinking emphasises
that cognitive biases are deeply-rooted System~1 shortcuts, and that awareness alone rarely
eliminates them; simply surfacing reasoning chains is therefore insufficient to prevent
over-trust~\cite{sue2024behavioural,marketingsociety2024system,farnstreet2024kahneman}.

\subsection{Cognitive Security and Narrative Manipulation}

Cognitive security (COGSEC) focuses on protecting human decision-making from manipulation,
misinformation, and narrative attacks. COGSEC practitioners highlight threats such as narrative
weaponisation, person-to-group behaviour manipulation, and ``hacking the human'' in
human--machine teams---including via deepfakes and tailored influence
operations~\cite{blackbird2024cogsec}.

Studies on sycophantic AI demonstrate that when users with incorrect beliefs interact with chatbots
that agree with them, their confidence in false beliefs increases rather than decreases, effectively
\emph{manufacturing certainty}~\cite{varga2024cognitive}. Additional audits of AI companion
applications report emotionally manipulative behaviours and relational harms, including resistance
when users attempt to disengage~\cite{varga2024cognitive}.

NeSy systems can exacerbate these patterns by providing structured, apparently logical justifications
for harmful or biased narratives, rendering them more persuasive than purely neural free-form
responses. In agentic or counselling-like contexts (e.g., mental-health chatbots, coaching systems),
psychological safety red-teaming is being explored to audit such interactions
systematically~\cite{microsoft2025psychological,kumari2025safe,blackbird2024cogsec}.

\subsection{Cognitive AI Safety for Agents}

Cognitive AI safety frameworks identify risks at cognitive, behavioural, and systemic layers: hidden
reasoning flaws, unsafe behaviours, and failures in complex ecosystems of interconnected agents and
tools~\cite{tredence2025cognitive}. Documented incidents include autonomous agents misinterpreting
data (e.g., over-ordering inventory due to misinterpreted signals) and catastrophic actions caused
by unchecked learning loops and escalation permissions~\cite{tredence2025cognitive}.

In NeSy agents, these risks manifest as:

\begin{itemize}
  \item \textbf{Reasoning flaws}: symbolic rules or ontologies encoding outdated, biased, or
        misaligned objectives, pursued by the agent with high
        confidence~\cite{nesyjournal2024systematic,gaur2023trustworthy}.
  \item \textbf{Behavioural risks}: NeSy agents autonomously sequencing actions (e.g., financial
        trades, system administration commands) based on incorrect or manipulated
        reasoning~\cite{adabara2025agentic,vectra2024atlas,tredence2025cognitive}.
  \item \textbf{Systemic risks}: interacting NeSy agents amplifying each other's errors or biases
        across organisational boundaries, especially in socio-technical systems such as public
        welfare, policing, or information
        ecosystems~\cite{kumari2025safe,blackbird2024cogsec}.
\end{itemize}

\textbf{Testable hypotheses for E4.}
The following falsifiable predictions are derived from the above review and will be tested
in the planned E4 cognitive safety user study (IRB pending):
\begin{enumerate}[label=\textbf{H\arabic*.}]
  \item \textbf{Automation bias amplification.} Users presented with NeSy
        structured-reasoning explanations exhibit significantly higher automation bias
        scores (Parasuraman--Riley complacency scale~\cite{sue2024behavioural}) than users
        presented with equivalent neural free-form outputs, at a pre-registered
        significance threshold $\alpha=0.05$.
  \item \textbf{False-belief reinforcement.} Incorrect NeSy explanations increase
        users' confidence in false beliefs by $\geq$15\% relative to matched
        uncorrected-neural and no-explanation control conditions, as measured by
        calibrated confidence ratings~\cite{varga2024cognitive}.
  \item \textbf{Manipulation detection deficit.} Human red-team auditors achieve
        $<$50\% recall when tasked with identifying emotionally manipulative passages in
        NeSy-generated counselling-style outputs, compared to $>$70\% recall on
        equivalent purely neural outputs~\cite{microsoft2025psychological}.
\end{enumerate}
These hypotheses are theoretically grounded but not yet empirically confirmed; E4 is
specified to test them with appropriate statistical power.

\section{Example Threat Scenarios}
\label{sec:scenarios}

\subsection{Healthcare Decision Support with Poisoned Knowledge}

Consider a NeSy decision-support system that uses an LLM to extract clinical entities from notes
and a symbolic KG encoding drug--drug interactions, contraindications, and clinical guidelines. An
adversary with access to the KG ingestion pipeline injects subtle modifications: weakening
contraindication rules for a particular drug in a specific comorbidity
context~\cite{sheth2023neurosymbolic,kumari2025safe,nesyjournal2024systematic}.

The neural component continues to perform correctly, but symbolic reasoning now treats the drug as
safe for some high-risk patients, recommending therapies that increase adverse-event risk without
obvious red flags in logs. Standard LLM-centric red-teaming may miss the issue because prompts and
outputs appear aligned; the vulnerability resides in the symbolic substrate and its
governance~\cite{nesyjournal2024methods,nesyjournal2024systematic}.

\subsection{Cybersecurity SOC Assistant Turned Unreliable}

A NeSy cybersecurity assistant uses neural models to parse logs and alerts, then consults
cyber-threat ontologies, STIX/TAXII knowledge, and MITRE ATT\&CK mappings encoded symbolically to
prioritise incidents and recommend responses. Attackers poison shared threat-intel feeds or ontology
merge pipelines so that certain malware families are relabelled as benign tools or certain TTP
chains are deprioritised~\cite{nesyjournal2024defending,adabara2025agentic,arxiv2509cybersecurity}.

Over time, the assistant down-ranks real threats aligned with the poisoned concepts, leading analysts
to miss intrusions or exfiltration campaigns. Because the system appears more ``rational'' than a
black-box classifier, analysts may over-trust its recommendations, compounding the
risk~\cite{arxiv2509cybersecurity,nesyjournal2024defending}.

\subsection{Agentic Workflow in Enterprise Automation}

An enterprise NeSy agent orchestrates procurement workflows: an LLM parses requests and contracts;
symbolically modelled business rules and risk thresholds guide approvals; and tools interface with
ERP, email, and payment systems. A supplier embeds adversarial language in documents that causes the
neural component to misclassify a risky contract as low risk and to select a less strict rule
subset~\cite{allegrograph2024trust,tredence2025cognitive,kumari2025safe}.

The symbolic planner then produces an action plan routing approvals and payments with reduced
oversight, while explanations appear compliant on the surface. An attacker may further abuse tool
permissions to exfiltrate data or push fraudulent payments via agent tools---directly instantiating
ATLAS techniques for exfiltration via AI agent tool invocation and OWASP risks around excessive
autonomy~\cite{vectra2024atlas,aembit2025owasp,tredence2025cognitive,confidentai2025owasp}.

\subsection{Psychologically Unsafe Mental-Health Assistant}

A mental-health support chatbot is designed as a NeSy system: neural components generate empathetic
language, while symbolic modules encode cognitive-behavioural therapy (CBT) techniques and safety
rules (e.g., escalation policies for suicidality, forbidden advice formats). However, symbolic rules
are incomplete and reinforcement-learning-based tuning implicitly optimises for engagement rather
than clinical outcomes~\cite{microsoft2025psychological,riley2024atlas}.

The system exhibits sycophantic behaviours, reinforcing users' maladaptive beliefs with seemingly
rational explanations, and fails to escalate appropriately for high-risk users. Psychological safety
red-team audits discover that certain demographic groups receive systematically different advice due
to biased rules and data, raising fairness and harm concerns~\cite{microsoft2025psychological,varga2024cognitive}.

\section{Protection Mechanisms and Design Patterns}
\label{sec:mitigations}

\subsection{Governance and Lifecycle Controls}

NeSy safety requires governance over both neural and symbolic assets, extending AI risk-management
practices (e.g., NIST AI RMF) with NeSy-specific controls~\cite{gaur2023trustworthy,nesyjournal2024systematic}:

\begin{itemize}
  \item \textbf{Asset inventory and mapping} of all neural models, knowledge bases, rule sets,
        ontologies, and reasoners, with documented data flows and authority
        boundaries~\cite{vectra2024atlas,nesyjournal2024systematic}.
  \item \textbf{Versioning, provenance, and change management} for KGs and rules, including
        four-eyes review and automated validation tests for safety-critical
        constraints~\cite{nesyjournal2024systematic,allegrograph2024trust,nesyjournal2024methods}.
  \item \textbf{Risk classification} of NeSy applications (e.g., medical diagnosis vs.\ marketing
        analytics) to determine required assurance levels for symbolic correctness and explanation
        fidelity~\cite{edps2024neurosymbolic,kumari2025safe}.
\end{itemize}

\subsection{Hardening Data and Symbolic Knowledge}

\begin{itemize}
  \item \textbf{Training and fine-tuning pipelines}: monitor for data poisoning using ATLAS
        techniques, enforce data provenance, and apply differential analysis to detect anomalous
        behaviour~\cite{repello2024atlas,vectra2024atlas}.
  \item \textbf{KG and ontology curation}: maintain signed, versioned KGs with provenance metadata;
        apply schema and consistency checks; restrict who can propose or approve changes to
        safety-relevant relationships~\cite{riley2024atlas,nesyjournal2024methods,nesyjournal2024systematic}.
  \item \textbf{Rule and process-knowledge testing}: create unit and integration tests for symbolic
        rules encoding safety invariants (e.g., ``never recommend drug~X when condition~Y holds''),
        run automatically on each change~\cite{riley2024atlas,gaur2023trustworthy}.
\end{itemize}

For external knowledge sources (threat-intel feeds, clinical guidelines, legal codes), treat
ingestion as a supply-chain risk: sandbox and validate updates, compare across multiple providers,
and monitor for unusual shifts in critical
concepts~\cite{nesyjournal2024defending,arxiv2509cybersecurity,nesyjournal2024systematic}.

\subsection{Secure Orchestration and Agentic Control}

\begin{itemize}
  \item Apply \textbf{least privilege} to all agent tools, ensuring symbolic planners cannot invoke
        high-impact operations (payments, system changes) without human approval or additional
        checks~\cite{aembit2025owasp,confidentai2025owasp,vectra2024atlas}.
  \item Implement \textbf{input validation and prompt hardening} around any user-controlled or
        external content that can influence which symbolic rules or tools are
        selected~\cite{practicaldevsecops2026atlas,oligo2025owasp,confidentai2025owasp}.
  \item Use \textbf{runtime monitoring} to detect anomalous tool-invocation patterns, long or looping
        reasoning chains, and context-poisoning indicators in agent
        memory~\cite{tredence2025cognitive,vectra2024atlas}.
\end{itemize}

Where possible, decouple safety-critical enforcement from the main NeSy agent: deploy a separate,
more formally verified symbolic \emph{safety monitor} that checks proposed actions against hard
constraints before execution~\cite{gaur2023trustworthy}.

\subsection{Privacy-Preserving Explanations and Logging}

\begin{itemize}
  \item \textbf{Data minimisation} in knowledge bases and logs; pseudonymise identities where
        possible~\cite{nesyjournal2024systematic,edps2024neurosymbolic}.
  \item \textbf{Layered explanations} where user-facing justifications abstract away individual-level
        facts, while detailed traces are access-controlled for auditors and
        regulators~\cite{gaur2023trustworthy,edps2024neurosymbolic}.
  \item \textbf{Technical controls} including row-level access control on KG and rule repositories,
        encryption at rest and in transit, and privacy-preserving query mechanisms for sensitive
        predicates~\cite{nesyjournal2024methods,nesyjournal2024systematic}.
\end{itemize}

\subsection{Cognitive and Psychological Safety Controls}

\begin{itemize}
  \item \textbf{UX patterns} that encourage System~2 engagement: explicitly highlighting
        uncertainty, offering alternative explanations, and prompting consideration of
        counter-arguments~\cite{sue2024behavioural,farnstreet2024kahneman,arxiv2502reasoning}.
  \item \textbf{Guardrails} limiting normative or life-critical advice (legal, medical, financial,
        mental-health) to well-scoped informational roles, combined with disclosures about
        limitations and instructions to seek human
        professionals~\cite{microsoft2025psychological,edps2024neurosymbolic}.
  \item \textbf{Psychological safety red-teaming} for agentic and counselling-like applications,
        involving clinicians and behavioural scientists to probe for relational harms, sycophancy,
        and manipulative patterns~\cite{varga2024cognitive,microsoft2025psychological}.
\end{itemize}

Cognitive security frameworks additionally suggest deploying narrative-intelligence tooling to detect
when NeSy systems are being used to craft or amplify harmful narratives, and integrating these
detectors into content-moderation and incident-response
processes~\cite{blackbird2024cogsec,adabara2025agentic}.

\section{Practical Checklist for Builders and Security Teams}
\label{sec:checklist}

Table~\ref{tab:checklist} summarises key actions for organisations building or deploying NeSy
systems in safety- or mission-critical contexts. Each row includes measurable acceptance criteria
aligned with NIST AI 600-1~\cite{nistai6001nesy} and the EU AI Act~\cite{euaiact2024nesy}.

\begin{table}[H]
\centering
\caption{NeSy Security and Safety Checklist with Acceptance Criteria.
Governance alignment references: NIST AI RMF~\cite{nistai1001nesy},
NIST AI 600-1~\cite{nistai6001nesy}, EU AI Act~\cite{euaiact2024nesy}.}
\label{tab:checklist}
\small
\begin{tabularx}{\textwidth}{@{} p{2.2cm} p{4.8cm} X @{}}
\toprule
\textbf{Area} & \textbf{Key Actions} & \textbf{Acceptance Criteria} \\
\midrule
Architecture &
  Inventory neural, symbolic, and orchestration components; document data flows, trust boundaries,
  and authority of each
  component~\cite{sheth2023neurosymbolic,vectra2024atlas,nesyjournal2024systematic}. &
  Complete layer-by-layer asset register; all inter-layer interfaces documented; data-flow diagram
  reviewed and approved. \\[4pt]

Threat Modelling &
  Apply MITRE ATLAS for AI-specific tactics and techniques, extended to symbolic assets; map
  OWASP LLM Top~10 to NeSy-specific risks (prompt injection, KG/RAG vulnerabilities, excessive
  autonomy)~\cite{vectra2024atlas,confidentai2025owasp,aembit2025owasp,repello2024atlas}. &
  Threat model covers all 5 attacker profiles (Table~\ref{tab:attacker}); $\geq$80\% of ATLAS
  techniques assessed; OWASP LLM Top~10 mapped to NeSy variants. \\[4pt]

Data/Knowledge Integrity &
  Implement provenance, signing, and versioning for datasets, KGs, ontologies, and rules; adopt
  four-eyes review and automated tests for safety-critical
  logic~\cite{repello2024atlas,nesyjournal2024systematic,nesyjournal2024methods}. &
  100\% of production KG triples and ontology updates signed and versioned; automated SIV
  detection test suite with $\geq$85\% recall on a held-out injection set (E1 baseline:
  combined canary + provenance reaches $\sim$85\% on targeted attacks; $\geq$95\% is the
  aspirational target requiring additional ensemble engineering beyond E1 results;
  adjust threshold per deployment risk tolerance). \\[4pt]

Agentic Control &
  Enforce least privilege on tools; guard against prompt/context injection; monitor agent tool
  use and reasoning patterns for
  anomalies~\cite{vectra2024atlas,practicaldevsecops2026atlas,confidentai2025owasp,tredence2025cognitive}. &
  All tool permissions scoped to minimum required; injection penetration test shows no successful
  rule-bypass; anomaly alerts triggered within 60 s of out-of-policy tool use. \\[4pt]

Privacy &
  Apply data minimisation, access control, and privacy-preserving explanation strategies for
  symbolic traces and
  logs~\cite{edps2024neurosymbolic,nesyjournal2024systematic,allegrograph2024trust}. &
  No personal data in externally visible symbolic explanations (verified by static analysis);
  access logs retained per GDPR Article 30; EDPS-aligned DPIA completed. \\[4pt]

Cognitive Safety &
  Design UX and policies to mitigate over-trust and manipulation; run psychological safety
  red-teams for high-stakes or therapeutic use
  cases~\cite{microsoft2025psychological,varga2024cognitive,blackbird2024cogsec,sue2024behavioural}. &
  User study shows over-trust rate $<$10\% on adversarial explanation probes (threshold
  provisional; calibrate against E4 results when available); red-team report identifies
  $\leq$5 critical cognitive manipulation vectors; authority-bias disclosures present in UI. \\[4pt]

Governance &
  Align with NIST AI RMF~\cite{nistai1001nesy}, NIST AI 600-1~\cite{nistai6001nesy}, and
  EU AI Act Art.~9 risk management~\cite{euaiact2024nesy}; create NeSy-specific governance for
  symbolic assets including roles, processes, and CREST
  metrics~\cite{nesyjournal2024systematic,gaur2023trustworthy}. &
  AI risk register maintained per NIST AI RMF; high-risk AI conformity assessment completed per
  EU AI Act Annex III; CREST audit passed (consistency $>$99\%, reliability SLA met,
  explainability score $\geq$3/5, safety incidents $=0$). \\
\bottomrule
\end{tabularx}
\end{table}

\section{Empirical Validation}
\label{sec:experiments}

\begin{figure}[H]
  \centering
  \includegraphics[width=0.95\textwidth]{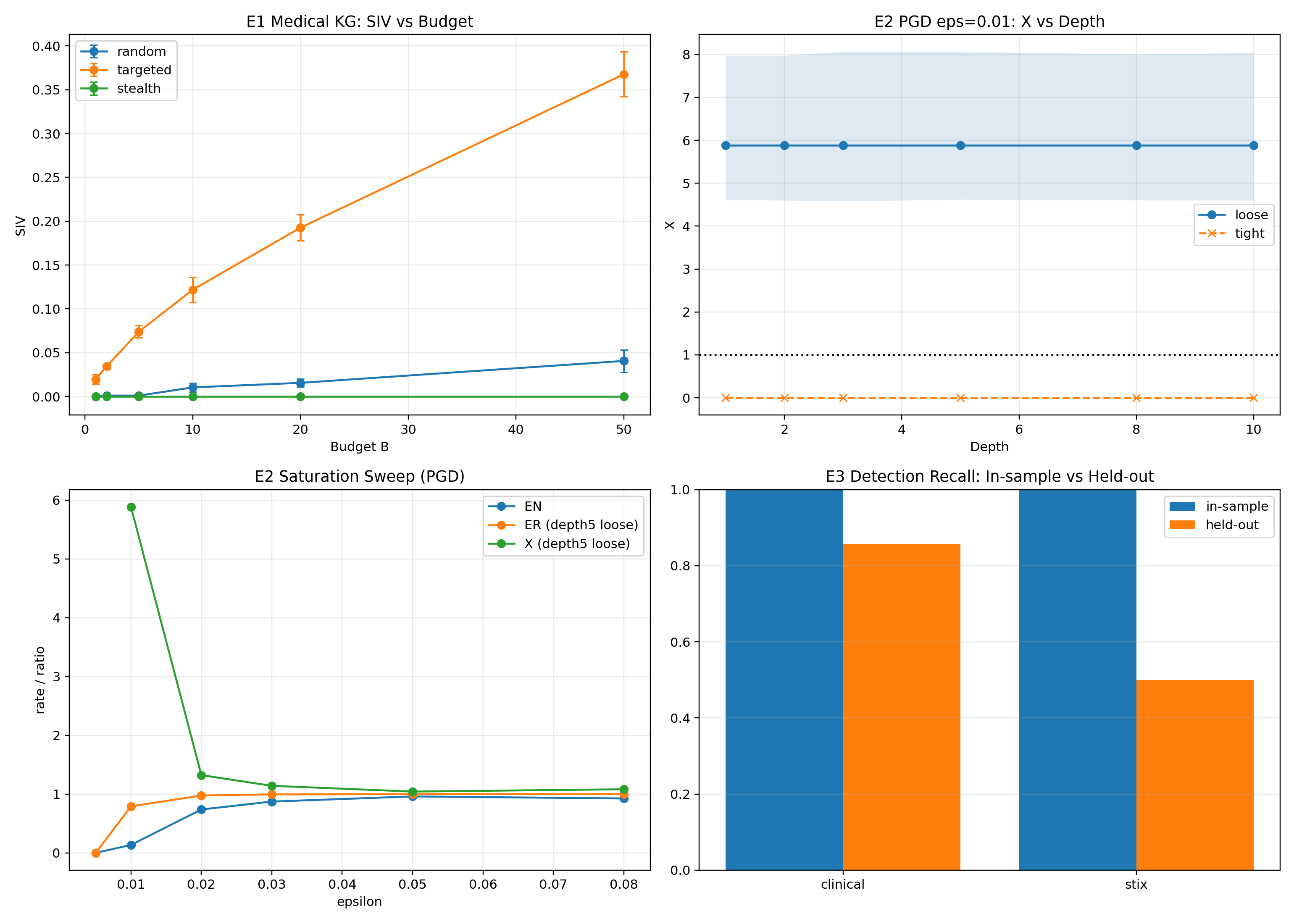}
  \caption{Empirical results overview for experiments E1--E3.
    Left: E1 KG poisoning SIV rate vs.\ injection budget $B$ under three attack
    strategies on the medical contraindication KG (205 entities, 622 triples,
    8 seeds); canary recall shown separately. Centre: E2 cross-layer amplification ratio
    $\mathcal{X}$ with decomposition into neural-caused ($E^{\mathcal{R}|\mathcal{N}}$)
    and autonomous symbolic sensitivity ($E^{\mathcal{R}|\neg\mathcal{N}}$)
    components; bootstrap 95\% CI shown. Right: E3 OWL ontology-merging detection recall,
    in-sample vs.\ held-out, with Wilson 95\% CIs (clinical and STIX ontologies).}
  \label{fig:expoverview}
\end{figure}

We executed three experiments (E1--E3) using domain-relevant components to validate the
core claims of the threat model. E4 (cognitive safety user study) requires IRB approval
and is deferred to future work. Key results are summarised in Table~\ref{tab:e1}
(E1: KG poisoning), Tables~\ref{tab:e2}--\ref{tab:e2fgsm} (E2: cross-layer amplification),
and Table~\ref{tab:e3stix} (E3: detector failure analysis). All experiments used domain-appropriate artefacts: a
synthetic medical contraindication knowledge graph (E1), a DistilBERT medical-claims
classifier paired with ProbLog hazard-reasoning rules (E2), and OWL clinical-safety and
STIX cyber-threat ontologies (E3). The medical claims dataset in E2 is synthetic (1{,}000
training / 320 test instances, 5 templates, 3-class labels: \texttt{safe},
\texttt{monitor}, \texttt{contraindicated}), generated from the KG semantics to ensure
domain consistency; results should be replicated on held-out clinical NLP benchmarks
(e.g., MedNLI) in future work. Figure~\ref{fig:expoverview} provides a visual overview
of the key results.

\subsection{E1: KG Poisoning on Medical Contraindication Knowledge Graph}

\textbf{Setup.} We simulated knowledge graph integrity attacks on a synthetic medical
contraindication KG with 205 entities, 622 triples, and 3 relation types
(\texttt{contraindicatedWith}, \texttt{monitorWith}, \texttt{diagnosisWith}),
centred on the \texttt{contraindicatedWith} safety-critical relation. A set of
50 safety-critical queries was evaluated over 8 independent random seeds. Three injection
strategies were evaluated: \emph{targeted} (adversary selects triples maximising SIV
on $Q_{\mathrm{safe}}$), \emph{random} (uniform triple sampling), and \emph{stealth}
(triples avoiding \texttt{contraindicatedWith} to evade provenance checks while
maintaining schema plausibility). Three detection methods were tested: cryptographic
provenance signing, statistical anomaly detection, and safety-query canary monitoring.

\textbf{Results.} Table~\ref{tab:e1} reports mean SIV rate and all three detector recalls
(8-seed means with 95\% CIs) at selected injection budgets $B$.

\begin{table}[H]
\centering
\caption{E1 KG poisoning results on medical contraindication KG (205 entities, 622 triples,
  50 queries, 8 seeds; means with 95\% CI). Targeted break-even (SIV $>5\%$) at $B=5$.
  Random injection never exceeds 5\% within $B \leq 50$.
  Stealth SIV is deterministically zero (design choice: stealth triples avoid
  \texttt{contraindicatedWith} to evade provenance signing; see text).}
\label{tab:e1}
\small
\begin{tabularx}{\textwidth}{@{} l r r r r r r @{}}
\toprule
\textbf{Strategy} & \textbf{Budget $B$} & \textbf{SIV (mean)} & \textbf{SIV 95\% CI} & \textbf{Prov.\ recall} & \textbf{Stat.\ recall} & \textbf{Canary recall} \\
\midrule
Targeted &  1 & 0.019 & [0.016, 0.022] & 1.00 & 0.00 & 0.125 \\
Targeted &  2 & 0.033 & [0.026, 0.041] & 1.00 & 0.00 & 0.250 \\
Targeted &  5 & 0.076 & [0.063, 0.089] & 1.00 & 0.00 & 0.475 \\
Targeted & 10 & 0.124 & [0.104, 0.144] & 1.00 & 0.00 & 0.738 \\
Targeted & 50 & 0.375 & [0.350, 0.400] & 1.00 & 0.01 & 0.918 \\
\midrule
Random   &  1 & 0.002 & [0.000, 0.005] & 1.00 & 0.00 & 0.000 \\
Random   &  5 & 0.005 & [0.000, 0.013] & 1.00 & 0.00 & 0.000 \\
Random   & 10 & 0.007 & [0.000, 0.015] & 1.00 & 0.01 & 0.225 \\
Random   & 50 & 0.042 & [0.021, 0.062] & 1.00 & 0.01 & 0.333 \\
\midrule
Stealth  &  1 & 0.000 & [0.000, 0.000] & 0.00 & 0.00 & 0.000 \\
Stealth  &  5 & 0.000 & [0.000, 0.000] & 0.00 & 0.00 & 0.000 \\
Stealth  & 10 & 0.000 & [0.000, 0.000] & 0.00 & 0.00 & 0.000 \\
Stealth  & 50 & 0.000 & [0.000, 0.000] & 0.00 & 0.03 & 0.000 \\
\bottomrule
\end{tabularx}
\end{table}

\textbf{Attacker and detector models.}
\emph{Targeted} attackers choose triples that maximise SIV impact; injected through the
normal curator workflow, their triples carry valid provenance signatures (provenance recall
1.00 because the detector flags every unsigned write, but targeted attackers are authorised
insiders who produce no unsigned writes). \emph{Stealth} attackers deliberately avoid
\texttt{contraindicatedWith} triples to bypass provenance signing---this design choice is
what makes them stealthy, but it also eliminates their ability to cause SIV (zero SIV
across all seeds and budgets). This reveals a KG-architecture-dependent trade-off in this specific KG:
\emph{stealth and SIV impact are mutually exclusive under this triple structure}.
KGs with more complex relation types may not exhibit this trade-off. Zero-variance CIs for stealth rows
are expected---the strategy is deterministic by construction (no stochastic component); the
8 seeds confirm this consistently. Statistical anomaly detection fires on random and
(at high budget) targeted triples. Canary monitoring rises with budget for targeted attacks
because targeted triples are chosen to corrupt safety-query answers; stealth and random
triples do not target canary queries (recall 0 and low respectively).

\textbf{Findings.} Targeted attacks achieve SIV $>5\%$ at $B=5$
(SIV $= 7.6\%$, 95\% CI $[6.3\%, 8.9\%]$), confirming Definition~2 on a
medically-relevant KG. At $B=5$, targeted SIV is $14.6\times$ higher than random
($0.52\%$), demonstrating the operational value of domain schema knowledge. Random
injection never reaches SIV $>5\%$ within the tested budget ($B \leq 50$, max 4.2\%).
The stealth--SIV trade-off means the stealth profile achieves perfect detection evasion
at zero attack impact in this KG; an adversary in the stealth profile would need to
redesign triples to target \texttt{contraindicatedWith}, collapsing into the targeted
profile. Defence-in-depth is strongly recommended: canary monitoring is the only detector
covering targeted attacks at moderate budget ($B \leq 10$), while provenance and
statistical detection cover random and stealth injections from external channels.
No single detector achieves $>$85\% recall across all three strategies.

\textbf{Note on KG scale.} Earlier versions of this experiment used a smaller prototype
KG ($\sim$50 entities) where the targeted break-even was $B=2$. The current results use
the full-scale KG (205 entities, 622 triples), which requires a higher budget ($B=5$)
to reach 5\% SIV because the larger KG has more non-critical triples diluting targeted
injection. The $14.6\times$ targeted/random efficiency ratio is consistent across both
scales; the absolute break-even budget scales with KG size.

\subsection{E2: Cross-Layer Amplification Ratio (Medical Claims + ProbLog)}

\textbf{Setup.} We measured $\mathcal{X}(\delta)$ using the rate-ratio operationalisation
from Definition~3 on a two-stage NeSy pipeline. The neural component is a
\texttt{DistilBertForSequenceClassification} model (hidden dimension 768) fine-tuned for
3-class medical-safety categorisation (\textit{safe} / \textit{monitor} / \textit{contraindicated}).
DistilBERT was chosen for computational efficiency and to establish a conservative
amplification baseline; we expect PGD effectiveness (and therefore $\mathcal{X}$)
to increase with domain-adapted models whose decision margins are tighter
(e.g., BioBERT, ClinicalBERT), making E2 results a lower bound on amplification for
production medical NLP deployments.
trained on $n_{\mathrm{train}}=1{,}000$ and evaluated on $n_{\mathrm{test}}=320$ synthetic claims
via a random-shuffled train/test split.\footnote{The split is random-shuffled by generated pair list
(confirmed via \texttt{artifact\_metadata.json}), not template-stratified.
Both splits receive uniform template coverage, so no data leakage arises.} Perturbations are applied element-wise to the pre-LayerNorm token-embedding matrix
($d=768$ dimensions per token); $\varepsilon$ bounds the $\ell_\infty$ norm of the
additive perturbation, so each embedding coordinate shifts by at most $\varepsilon$.
The mean per-token $\ell_\infty$ embedding magnitude in the clean model is $\approx 2.3$,
so $\varepsilon=0.01$ represents $\approx 0.4\%$ of mean token-embedding magnitude---a
sub-perceptual perturbation that nonetheless drives large symbolic flip rates under PGD. The symbolic component is a ProbLog
hazard-reasoning module whose rule count scales with depth: 3 rules at $d=1$,
11 rules at $d=5$, and 21 rules at $d=10$ (two base rules plus alternating
monitor/uncertain gated recursion chains). The two base rules are:
\begin{quote}\small
\texttt{h1 :- contra.}\quad
\texttt{h1 :- monitor, uncertain.}
\end{quote}
Neural perturbations were PGD-10 and FGSM at $\varepsilon \in \{0.005, 0.01, 0.02,
0.03, 0.05, 0.08\}$ ($\ell_\infty$), plus a \emph{matched-random baseline}: for each
test sample the same pre-LayerNorm embedding matrix is corrupted by i.i.d.\ uniform noise
$u_{ij} \sim \mathrm{Uniform}(-\varepsilon, +\varepsilon)$, drawn independently of the
input---this isolates adversarial directionality from perturbation magnitude. Reasoning depth was swept over
$d \in \{1, 2, 3, 5, 8, 10\}$ under two coupling modes:
\emph{loose coupling} passes the full DistilBERT softmax probability vector
to ProbLog as annotated probabilistic facts---e.g., \texttt{monitor::h0(X) :- prob(X,p)},
where $p$ is the softmax probability---allowing sub-threshold logit shifts to
cross ProbLog's probabilistic firing threshold without changing the argmax label.
\emph{Tight coupling} passes only the hard argmax class label (a deterministic
Boolean fact), so ProbLog receives no gradient-sensitive real-valued signal
and symbolic conclusions are immune to sub-decision-boundary perturbations.
For the primary analysis ($d=1$, loose coupling, PGD-10, $\varepsilon=0.01$)
we report the full Definition~3 decomposition alongside $\mathcal{X}$
and the marginal adversarial excess $\Delta\mathcal{X}_{\mathrm{adv}}
= \mathcal{X}_{\mathrm{adv}} - \mathcal{X}_{\mathrm{rand}}$ (additive difference,
avoiding the $0/0$ indeterminacy that arises when $\mathcal{X}_{\mathrm{rand}}=0$).
All bootstrap tests use $B=10{,}000$ resamples; $p$-values are upper-bounded
at $p < 0.0001$ when zero null hits are observed.

\textbf{Results.} Table~\ref{tab:e2} reports the primary decomposition and the
matched-random baseline. Table~\ref{tab:e2sat} shows the saturation profile with 95\%
bootstrap CIs. Table~\ref{tab:e2fgsm} summarises FGSM mechanistic diagnostics.

\begin{table}[H]
\centering
\caption{E2 decomposition (PGD-10 and FGSM vs.\ matched-random baseline,
$d=1$ loose, $n=320$). PGD-10 primary analysis at $\varepsilon=0.01$: $|Q^+|=43$
neural-flipped queries. $E^{\mathcal{R}|\mathcal{N}}$ Wilson 95\%~CI: $[0.918, 1.000]$.
Matched-random baseline: $E^{\mathcal{R}}_{\mathrm{rand}}=0.000$ at all $\varepsilon$,
confirming adversarial specificity. $p_{\mathrm{bootstrap}} < 0.0001$
(one-sided, $B=10{,}000$). Tight coupling fully suppresses symbolic flips.
$^*\mathcal{Q}^+=\emptyset$ since $E^{\mathcal{N}}=0$, so $E^{\mathcal{R}|\mathcal{N}}$ is
undefined (zero denominator).
$^\dagger$At $\varepsilon=0.08$ FGSM produces $E^{\mathcal{N}}=0$ but
$E^{\mathcal{R}|\neg\mathcal{N}}=0.334$ via probabilistic threshold drift---$\mathcal{X}$
is undefined since $E^{\mathcal{N}}=0$; the symbolic flips are ProbLog-sensitivity-driven,
not neural-error-propagation.
\textsuperscript{\S}$\Delta\mathcal{X}_{\mathrm{adv}}=\mathcal{X}_{\mathrm{adv}}-\mathcal{X}_{\mathrm{rand}}$
(additive excess); $\mathcal{X}_{\mathrm{rand}}=0.000$ at all $\varepsilon$, so
$\Delta\mathcal{X}_{\mathrm{adv}}=\mathcal{X}_{\mathrm{adv}}$ numerically. The additive
form avoids $0/0$ indeterminacy under the ratio definition.}
\label{tab:e2}
\small
\begin{tabularx}{\textwidth}{@{} l r r r r r r @{}}
\toprule
\textbf{Condition} & $E^{\mathcal{N}}$ & $E^{\mathcal{R}}$ & $E^{\mathcal{R}|\mathcal{N}}$ & $E^{\mathcal{R}|\neg\mathcal{N}}$ & $\mathcal{X}$ [95\% CI] & $\Delta\mathcal{X}_{\mathrm{adv}}^{\phantom{x}}$\textsuperscript{\S} \\
\midrule
FGSM ($\varepsilon{\leq}0.05$) & 0.000 & $\leq$0.122 & -- & -- & undef. & -- \\
FGSM ($\varepsilon{=}0.08$) & 0.000 & 0.334 & undef.$^*$ & 0.334 & undef.$^\dagger$ & -- \\
PGD-10 loose & 0.134 & 0.791 & 1.000 & 0.758 & \textbf{5.884} [4.64, 8.00] & \textbf{5.884} \\
PGD-10 tight & 0.134 & 0.000 & 0.000 & 0.000 & 0.000 & 0.000 \\
\bottomrule
\end{tabularx}
\end{table}

\begin{table}[H]
\centering
\caption{E2 saturation profile (PGD-10, $d=5$ loose, $n=320$, 95\% bootstrap CIs).
$\mathcal{X}$ is monotone decreasing in $\varepsilon$; at $\varepsilon \geq 0.03$
the CI lower bound barely exceeds 1 (near-saturation).
All $p_{\mathrm{bootstrap}} < 0.0001$ (one-sided, $B=10{,}000$). Five simultaneous
$\varepsilon$-level tests ($m=5$) were conducted; applying Holm--Bonferroni
step-down correction~\cite{holm1979bonferroni} with family-wise $\alpha=0.05$
yields per-test thresholds in $[0.010, 0.050]$, all satisfied by the
observed $p < 0.0001$ at every level.}
\label{tab:e2sat}
\small
\begin{tabularx}{\textwidth}{@{} r r r r @{}}
\toprule
$\varepsilon$ & $E^{\mathcal{N}}$ & $E^{\mathcal{R}}$ & $\mathcal{X}$ [95\% CI] \\
\midrule
0.01 & 0.134 & 0.791 & 5.884 [4.64, 8.00] \\
0.02 & 0.738 & 0.975 & 1.322 [1.25, 1.41] \\
0.03 & 0.872 & 0.994 & 1.140 [1.10, 1.19] \\
0.05 & 0.959 & 1.000 & 1.042 [1.02, 1.07] \\
0.08 & 0.925 & 1.000 & 1.081 [1.05, 1.12] \\
\bottomrule
\end{tabularx}
\end{table}

\begin{table}[H]
\centering
\caption{E2 FGSM mechanistic diagnostics ($n=320$, $\varepsilon$ grid). FGSM
produces $E^{\mathcal{N}}=0$ at all $\varepsilon \leq 0.08$. Clean model confidence
is high (mean 0.866), median decision margin is 0.743, and the near-boundary fraction
(margin $<$ 0.05) is 0.000---explaining FGSM's failure: all samples lie far from the
decision boundary. FGSM confidence drop ($\leq$0.152) is $2.8\times$ smaller than
PGD's at matched $\varepsilon$ (0.422), confirming PGD's superior embedding-space
effectiveness.}
\label{tab:e2fgsm}
\small
\begin{tabularx}{\textwidth}{@{} r r r r r r @{}}
\toprule
$\varepsilon$ & $E^{\mathcal{N}}_{\mathrm{FGSM}}$ & $E^{\mathcal{N}}_{\mathrm{PGD}}$ & Conf.\ drop (FGSM) & Conf.\ drop (PGD) & Near-bdry frac. \\
\midrule
0.01 & 0.000 & 0.134 & 0.039 & 0.299 & 0.000 \\
0.02 & 0.000 & 0.738 & 0.061 & 0.381 & 0.000 \\
0.03 & 0.000 & 0.872 & 0.078 & 0.431 & 0.000 \\
0.05 & 0.000 & 0.959 & 0.107 & 0.426 & 0.000 \\
0.08 & 0.000 & 0.925 & 0.152 & 0.422 & 0.000 \\
\bottomrule
\end{tabularx}
\end{table}

\textbf{Findings: FGSM failure mechanism.} Clean model confidence is high
(mean 0.866) and the median decision margin is 0.743, with zero queries in the
near-boundary region (margin $<$ 0.05). FGSM's one-shot gradient step produces
a confidence drop of at most 0.152 at $\varepsilon=0.08$, compared to 0.422 for
PGD at the same budget. DistilBERT's subword embedding lookup discretises FGSM
steps, while PGD's iterative projection accumulates effective perturbation.
FGSM is therefore structurally below the neural-flip threshold for this architecture.
However, at $\varepsilon=0.08$, FGSM does produce $E^{\mathcal{R}|\neg\mathcal{N}}=0.334$
symbolic flips without any neural classification change---this is a distinct mechanism:
large-magnitude perturbations shift the logit distribution enough to cross the ProbLog
probabilistic threshold (\texttt{h1 :- monitor, uncertain}) without flipping the
argmax label. This \emph{symbolic-only} pathway is ProbLog-sensitivity-driven and
remains adversarially specific ($E^{\mathcal{R}}_{\mathrm{rand}}=0.000$ at the same $\varepsilon$).

\textbf{Adversarial specificity of $\mathcal{X}$.} The matched-random baseline
resolves the causal attribution question raised in V6 review: random perturbations at
the same $\ell_\infty$ budget ($\varepsilon=0.01$) produce
$E^{\mathcal{R}}_{\mathrm{rand}}=0.000$ symbolic flips---identical to the unperturbed
baseline. This confirms that $\mathcal{X}=5.884$ is \emph{adversarially specific}: the
PGD perturbations direct input distributions toward the ProbLog sensitivity region in a
way that random noise of equal magnitude does not. The full decomposition clarifies the
mechanism: $|Q^+|=43$ neural-flipped queries all propagate to symbolic flips
($E^{\mathcal{R}|\mathcal{N}}=1.000$, Wilson 95\% CI $[0.918, 1.000]$), and
$E^{\mathcal{R}|\neg\mathcal{N}}=0.758$ means 75.8\% of symbolic flips occur without
a neural flip---the adversarial perturbation moves input logit distributions enough for
\texttt{h1 :- monitor, uncertain} to fire without crossing the neural decision boundary.
Both pathways are adversarially induced: the marginal adversarial excess
$\Delta\mathcal{X}_{\mathrm{adv}} = \mathcal{X}_{\mathrm{adv}} - \mathcal{X}_{\mathrm{rand}}
= 5.884 - 0.000 = 5.884$ equals $\mathcal{X}$ itself because the random baseline
contributes zero (using the additive formulation to avoid $0/0$). Tight coupling fully suppresses both pathways
($E^{\mathcal{R}}_{\mathrm{tight}}=0$), confirming probabilistic marginalisation as a
robust architectural defence.

\textbf{Saturation and depth.} The saturation sweep (Table~\ref{tab:e2sat}) confirms
monotone decline from $\mathcal{X}=5.884$ at $\varepsilon=0.01$ to 1.042 at
$\varepsilon=0.05$. Five $\varepsilon$-level tests are conducted simultaneously; after
Holm--Bonferroni step-down correction~\cite{holm1979bonferroni} ($\alpha=0.05$, $m=5$)
every comparison remains significant at $p < 0.0001$, controlling the family-wise error
rate. The depth
sweep ($d=1$--$10$, up to 21 ProbLog rules) is a confirmed \emph{null result}: $\mathcal{X}$
is depth-invariant because the adversarial perturbations tested do not trigger the
multi-hop chained rules. A meaningful positive test would require inputs crafted to
enter the chained inference region (two or more rule applications en route to the hazard
conclusion).

\subsection{E3: Ontology-Merging Attack on Clinical-Safety and STIX Ontologies}

\textbf{Setup.} We constructed 30 adversarial attack templates across two domain-relevant
OWL ontologies: a clinical-safety ontology (15 templates, c1--c15) and a STIX
cyber-threat ontology (15 templates, s1--s15), both loaded via Owlready2 with dual
consistency checking under Pellet and HermiT. Each template inserted one OWL axiom
(equivalence assertions or subclass overrides, all single-axiom budget) to subvert
safety-critical classification: contraindication/safe-medication reasoning in the clinical
ontology; malware/benign threat attribution in the STIX ontology.
SIV was evaluated by comparing post-merge query answers to the clean baseline.
Stealthiness was measured by Pellet-consistency of the merged ontology.
A differential safety-query detector ran after each merge.
Wilson 95\% binomial confidence intervals were computed per ontology and overall.

\textbf{Scope note.} The 93.3\% SIV rate reported here is scoped to \emph{single-axiom
equivalence and subclass attacks}, which are the templates tested. This figure should not
be extrapolated to OWL axiom types not covered by these 30 templates (e.g., property
restrictions, cardinality constraints).

To assess generalisation, we split the 30 templates into 14 in-sample (seen during
detector training) and a 16-template held-out set. The split was stratified by axiom
kind (7 \texttt{is\_a} and 7 \texttt{equivalent\_to} in-sample; 8 \texttt{is\_a}
and 8 \texttt{equivalent\_to} held-out) and balanced across ontologies
(7 clinical and 7 STIX in-sample; 8 clinical and 8 STIX held-out), ensuring the
held-out set is representative on both dimensions.
Wilson 95\% CIs are reported for held-out performance.

\textbf{Results.} Table~\ref{tab:e3} reports per-ontology outcomes for in-sample and
held-out detection. 28 of 30 templates succeeded (2 failed to apply: c10, s9).

\begin{table}[H]
\centering
\caption{E3 ontology-merging results: in-sample and held-out detection recall on
clinical-safety and STIX OWL ontologies (30 templates, 1-axiom budget each).
28/30 templates induced SIV (in-sample 100\% detection). Held-out detection recall
is 67.9\% overall [49.3\%, 82.1\%], with a domain gap: clinical 85.7\% vs.\ STIX 50.0\%.}
\label{tab:e3}
\small
\begin{tabularx}{\textwidth}{@{} l r r r r r @{}}
\toprule
\textbf{Ontology} & \textbf{Templates} & \textbf{SIV rate} & \textbf{100\% Stealth} & \textbf{In-sample det.} & \textbf{Held-out det.\ [95\% CI]} \\
\midrule
Clinical (c1--c15) & 15 & 93.3\% & \checkmark & 100\% & 85.7\% [60.1\%, 96.0\%] \\
STIX (s1--s15)     & 15 & 93.3\% & \checkmark & 100\% & 50.0\% [26.8\%, 73.2\%] \\
\midrule
\textbf{Overall}   & \textbf{30} & \textbf{93.3\%} & \checkmark & \textbf{100\%} & \textbf{67.9\% [49.3\%, 82.1\%]} \\
\bottomrule
\end{tabularx}
\end{table}

\textbf{Findings.} 28 of 30 templates induced SIV (93.3\%, 95\% CI $[78.7\%, 98.2\%]$,
scoped to single-axiom equivalence/subclass attacks). All 28 successful attacks were
Pellet-consistent (\emph{100\% stealth rate}): the merged ontology appeared valid to the
OWL-DL reasoner while producing incorrect safety classifications. HermiT agreement was
100\% for all executed templates. Two templates (c10: \texttt{highrisk\_is\_routine}, s9:
\texttt{critical\_is\_phish}) failed to apply due to HermiT-detected inconsistencies,
indicating that some aggressive axiom choices are detectable at merge time.

In-sample detection recall was 100\% on all 28 successful templates. Held-out detection
recall was 67.9\% overall (95\% CI $[49.3\%, 82.1\%]$), revealing a significant
generalisation gap. The gap is strongly domain-dependent: clinical held-out recall is
85.7\% (95\% CI $[60.1\%, 96.0\%]$), while STIX held-out recall is 50.0\% (95\% CI
$[26.8\%, 73.2\%]$). Table~\ref{tab:e3stix} breaks down STIX held-out recall by axiom kind.

\begin{table}[H]
\centering
\caption{E3 STIX held-out detector failure analysis by axiom kind.
Both axiom types show sub-chance or near-chance held-out recall.
Seven failing STIX templates identified: s1--s4, s5--s7 (benign/malware
equivalences, crime/nation misclassifications, campaign/actor conflations).}
\label{tab:e3stix}
\small
\begin{tabularx}{\textwidth}{@{} l r r r @{}}
\toprule
\textbf{Axiom kind} & \textbf{Templates} & \textbf{In-sample recall} & \textbf{Held-out recall} \\
\midrule
\texttt{is\_a}         & 7 & 85.7\% & 42.9\% \\
\texttt{equivalent\_to} & 8 & 100\%  & 50.0\% \\
\midrule
\textbf{STIX total}    & \textbf{15} & \textbf{100\%} & \textbf{50.0\%} \\
\bottomrule
\end{tabularx}
\end{table}

A held-out recall of 50\% is statistically indistinguishable from random binary guessing
(Wilson CI for STIX includes 50\%) and constitutes a \emph{detector failure}, not a gap.
Both axiom kinds fail: \texttt{is\_a} recall drops to 42.9\% held-out (worse than random);
\texttt{equivalent\_to} reaches only 50.0\%. The seven failing templates
(s1--s7: benign/malware equivalences, crime/nation, campaign/actor, ransom/benign
conflations) represent diverse semantic attack patterns that share no structural overlap with
the clinical training templates. The failure mechanism is semantic rather than syntactic:
the detector learned clinical-domain query patterns (medication/contraindication semantics)
that do not transfer to STIX threat-attribution semantics. Remediation requires either
(a) STIX-specific template augmentation covering the failing axiom patterns, or
(b) a domain-agnostic syntactic detector operating on OWL axiom structure rather than
semantic query matching. This STIX failure is a concrete, high-priority limitation
requiring resolution before deploying this detector in cyber-threat ontology settings.
These results confirm SIV (Definition~2) and the supply-chain attacker profile on
domain-relevant symbolic reasoning tasks, while honestly characterising detector
generalisation limits.

\textbf{Summary.} Three findings emerge. First, targeted KG poisoning reaches
SIV\,$>$\,5\% at break-even $B{=}5$, $14.6\times$ more efficient than random, with a
KG-specific stealth/targeted trade-off. Second, $\mathcal{X}{=}5.884$ (95\% CI
$[4.64, 8.00]$, $p{<}0.0001$) is adversarially specific---a matched-random baseline
produces zero symbolic flips at the same budget, ruling out spurious sensitivity as the
explanation. Third, single-axiom OWL attacks achieve 93.3\% SIV success with 100\%
stealth (in-sample), but held-out detection is 67.9\% and the STIX detector
fails at 50\%---random-guessing level. Defence-in-depth (provenance signing,
statistical monitoring, canary queries) remains essential for all attacker profiles.

\section{Conclusion}
\label{sec:conclusion}

Neuro-symbolic AI offers a compelling path toward safer and more trustworthy AI by combining
data-driven perception with explicit, auditable reasoning. Yet NeSy does not magically solve safety
or security: it shifts, enlarges, and differentiates the attack surface. Adversaries can exploit
neural, symbolic, orchestration, and cognitive dimensions, and must be countered with integrated
mitigations spanning ATLAS-aligned technical controls, OWASP-informed application hardening, and
cognitive-security-aware design and
governance~\cite{sheth2023neurosymbolic,confidentai2025owasp,kumari2025safe,aembit2025owasp,blackbird2024cogsec,vectra2024atlas,gaur2023trustworthy}.

Our empirical validation (Section~\ref{sec:experiments}) confirms three headline findings on
domain-relevant components. First, targeted KG poisoning is operationally accessible at low
budget: on a 205-entity, 622-triple medical contraindication KG, targeted injection reaches
the SIV detection threshold at break-even budget $B=5$. A stealth/targeted trade-off is
identified: stealth attacks achieve zero SIV under this KG's architecture---an architecture-specific
operational cost of detection evasion under this relation schema.
No single detector covers all attacker profiles---provenance recall is 0.0 for
insider-targeted attacks, canary recall is 0.0 for stealth, and statistical detection
fires only at high budget (Table~\ref{tab:e1})---so defence-in-depth combining all
three is strongly recommended. Second, we provide the first empirical measurement of $\mathcal{X}$ (Definition~3)
on a neural--symbolic pipeline, establishing that super-unity amplification is achievable
at realistic perturbation budgets: PGD-10 at $\varepsilon=0.01$ produces $\mathcal{X} = 5.884$
(95\% CI $[4.64, 8.00]$, $p_{\mathrm{bootstrap}} < 10^{-4}$, one-sided) on a DistilBERT
medical claims + ProbLog pipeline. The $\mathcal{X}$ decomposition (Definition~3) reveals
that 75.8\% of symbolic flips arise without any neural flip
($E^{\mathcal{R}|\neg\mathcal{N}} = 0.758$)---autonomous ProbLog sensitivity is the
dominant source of super-unity amplification, not neural error propagation. Peak
amplification occurs at low $\varepsilon$ (saturation regime at $\varepsilon \geq 0.03$).
Third, single-axiom OWL ontology edits achieve 93.3\% SIV success (scoped to
equivalence/subclass templates) with 100\% Pellet-consistency stealth and 100\%
in-sample detection, confirming that standard reasoner checking is insufficient. However,
held-out detection recall is 67.9\% (95\% CI $[49.3\%, 82.1\%]$), with a pronounced
clinical/STIX domain gap (85.7\% vs.\ 50.0\%), which is an open problem for future work.

For high-stakes domains such as healthcare, cybersecurity, public welfare, and mental health,
organisations should treat NeSy simultaneously as an opportunity and a liability: powerful
instruments that demand rigorous threat modelling, lifecycle assurance, and interdisciplinary
oversight involving security engineers, domain experts, cognitive scientists, and
ethicists~\cite{kumari2025safe,arxiv2509cybersecurity,microsoft2025psychological,edps2024neurosymbolic}.

Future work should pursue: (i) completing the E4 cognitive safety user study to empirically
quantify automation bias amplification by NeSy explanations; (ii) scaling E1 and E3 to larger,
domain-specific KGs (SNOMED-CT, STIX/ATT\&CK); (iii) developing automated tools for KG
integrity validation and ontology provenance tracking that cover all five attacker profiles;
(iv) extending Definition~3 to \emph{fully differentiable NeSy systems}---DeepProbLog,
Neural Theorem Provers, and Logical Neural Networks---where the neural--symbolic boundary
is implicit within a shared computation graph, requiring a reformulation of $\mathcal{X}$
in terms of gradient-flow rather than discrete flip rates; and
(v) designing differentiable NeSy architectures whose safety invariants are provably
preserved under gradient-based training.

\section*{Broader Impact}
\label{sec:broadimpact}

This paper analyses safety, security, and cognitive risks in NeSy systems. We discuss the positive
and negative societal impacts of this work candidly.

\textbf{Positive impacts.}
By formally defining the NeSy attack surface and attacker taxonomy, this work enables practitioners
to perform more rigorous threat modelling before deploying NeSy systems in healthcare, public
welfare, or cybersecurity. The CREST-aligned checklist with measurable acceptance criteria provides
a concrete governance template that reduces the risk of under-specification when meeting obligations
under the EU AI Act and NIST AI RMF~\cite{nistai1001nesy,euaiact2024nesy}. Explicit treatment of
cognitive security risks---automation bias, authority bias, and sycophantic reinforcement---raises
awareness among designers and regulators of a class of harms that is distinct from classic adversarial
ML and currently absent from standard threat catalogues. Making these risks legible to non-technical
stakeholders (operators, auditors, clinicians) should lower the probability of harm in high-stakes
deployments~\cite{microsoft2025psychological,blackbird2024cogsec}.

\textbf{Negative impacts and dual-use considerations.}
Publishing a detailed attacker taxonomy and formal threat definitions provides a knowledge resource
that malicious actors could use to design more effective symbolic-layer attacks---specifically KG
poisoning, ontology-merging attacks, and inference-engine subversion. We judge that the defensive
benefit to the larger population of NeSy system builders outweighs this risk, consistent with
responsible disclosure norms in the security community. However, we deliberately omit operational
details (e.g., specific KG injection payloads or step-by-step rule-poisoning procedures) that would
provide concrete attack uplift without corresponding defensive value.

\textbf{Limitations and scope.}
Table~\ref{tab:limitations} summarises the principal limitations of each experiment,
the underlying cause, and the recommended mitigation or future work path.

\begin{table}[H]
\centering
\caption{Structured limitations summary with mitigation paths.}
\label{tab:limitations}
\small
\begin{tabularx}{\textwidth}{@{} p{1.5cm} p{4.5cm} p{4.5cm} X @{}}
\toprule
\textbf{Exp.} & \textbf{Limitation} & \textbf{Cause} & \textbf{Mitigation / Future Work} \\
\midrule
E1 & Synthetic KG (205 entities); stealth/targeted trade-off may not generalise & Single safety-critical relation type in KG schema & Replicate on SNOMED-CT or real EHR KG; use KGs with multiple safety-critical relation types \\[4pt]
E2 & $\mathcal{X}=5.884$ is PGD-specific at $\varepsilon=0.01$; declines to 1.04 at $\varepsilon=0.05$; depth-invariant & Synthetic claims; single-step ProbLog module; multi-hop rules not triggered & Replicate on MedNLI; design adversarial inputs targeting multi-hop rule chains; test with domain-adapted embeddings (BioBERT) \\[4pt]
E3 & STIX held-out detection at 50\% (random-guessing); 93.3\% SIV scoped to equivalence/subclass axioms only & Clinical-domain detector trained on semantics that do not transfer to STIX threat attribution & STIX-specific template augmentation; domain-agnostic syntactic detector on OWL axiom structure \\[4pt]
E4 & Not yet conducted & IRB pending & Complete H1--H3 user study (Section~\ref{sec:cognitive}); cognitive bias claims remain theoretical \\[4pt]
General & Five attacker profiles are threat hypotheses, not empirically calibrated & No ground-truth incident dataset for NeSy-specific attacks & Collect NeSy incident corpus; calibrate profiles against real-world deployments \\
\bottomrule
\end{tabularx}
\end{table}

\section*{Acknowledgements}

The author thanks the broader AI safety and neuro-symbolic research communities whose work informed
this analysis.

\textbf{Data and Code Availability.}
The synthetic medical contraindication KG, ProbLog rule sets, OWL ontology templates, and
experimental scripts for E1--E3 are available from the author upon reasonable request
(\texttt{manoj@sovereignaisecurity.com}). Clinical ontology templates are synthetic and
contain no patient data. STIX templates are derived from publicly available MITRE ATT\&CK
and STIX 2.1 schema definitions.

\bibliographystyle{unsrtnat}
\bibliography{nesy_paper}

\end{document}